\documentclass[fleqn,usenatbib]{mnras}

\usepackage{newtxtext,newtxmath}
\usepackage{mathptmx}

\usepackage[T1]{fontenc}

\DeclareRobustCommand{\VAN}[3]{#2}
\let\VANthebibliography\thebibliography
\def\thebibliography{\DeclareRobustCommand{\VAN}[3]{##3}\VANthebibliography}

\usepackage{graphicx}	
\usepackage{amsmath}	

\newcommand{\hii}{\ion{H}{II}}
\newcommand{\HeI}{\ion{He}{I}}
\newcommand{\ArIII}{[\ion{Ar}{III}]} 
\newcommand{\SII}{[\ion{S}{II}]} 
\newcommand{\SIII}{[\ion{S}{III}]} 
  
\newcommand{\NI}{[\ion{N}{II}]} 
\newcommand{\NII}{[\ion{N}{II}]}  
\newcommand{\OIII}{[\ion{O}{III}]}  
\newcommand{\OII}{[\ion{O}{II}]}
\newcommand{\OI}{[\ion{O}{I}]}
\newcommand{\AV}{$A_V$} 
 
\newcommand{\Halpha}{H$\alpha$} 
\newcommand{\Hbeta}{H$\beta$} 
\newcommand{\nelec}{$n_e$}

\title[3D structure of S254-S258]{3D structure of \hii{} regions in the star-forming complex S254-S258}

\author[M. S. Kirsanova et al.]{
Maria S. Kirsanova,$^{1}$\thanks{E-mail: kirsanova@inasan.ru}
Alexei V. Moiseev,$^{2,1}$
Paul A. Boley$^{3,1}$
\\
$^{1}$Institute of Astronomy, Russian Academy of Sciences, 119017, 48 Pyatnitskaya Str., Moscow, Russia\\
$^{2}$Special Astrophysical Observatory, Russian Academy of Sciences, Nizhny Arkhyz 369167, Russia\\
$^{3}$Visiting astronomer, Laboratoire Lagrange, Universit\'e C\^ote d\textquoteright Azur, Observatoire de la C\^ote d\textquoteright Azur, CNRS, Boulevard de l\textquoteright Observatoire, CS 34229, 06304, Nice Cedex 4, France
}

\date{Accepted XXX. Received YYY; in original form ZZZ}

\pubyear{2023}

\begin{document}
\label{firstpage}
\pagerange{\pageref{firstpage}--\pageref{lastpage}}
\maketitle

\begin{abstract}
The S254-258 star-forming complex is a place of massive star formation where five OB-stars have created \hii{} regions, visible as optical nebulae, and disrupted the parental molecular gas. In this work, we study the 3D~structure of these \hii{} regions using optical spectroscopy and tunable-filter photometry with the 6-m and 1-m telescopes of the Special Astrophysical Observatory of the Russian Academy of Sciences. We construct maps of the optical extinction, and find that the \hii{} emission is attenuated by neutral material with $2 \leq A_V \leq 5$~mag. The typical electron density in S255, and S257 is $\approx 100$~cm$^{-3}$, with enhancements up to 200~cm$^{-3}$ in their borders, and up to 400~cm$^{-3}$ toward the dense molecular cloud between them, where active star formation is taking place. We show that either a model of a clumpy dense neutral shell, where UV~photons penetrate through and ionize the gas, or a stellar wind, can explain the shell-like structure of the ionized gas. S255 is surrounded by neutral material from all sides, but S257 is situated on the border of a molecular cloud and does not have dense front and rear walls. The compact \hii{} regions S256 and S258 are deeply embedded in the molecular clouds.
\end{abstract}




\section{Introduction}\label{sec:intro}

The appearance of \hii{} regions on optical images of the night sky is one of the most easily observed manifestations for the feedback from the young massive stars on the interstellar medium. Since at least the 1970s, it was observed that \hii{} regions can appear not only as sphere-like structures, as was previously considered in many theoretical works  \citep[e.g.][]{Stromgren_1939, Spitzer_1978, Hosokawa_2006, Kirsanova_2009, Bisbas_2015}. Instead, blister-like, bipolar, and irregular structures were found and modelled later by many authors \citep[e.g.][]{1979ApJ...233...85B, Tenorio-Tagle1979, 1990ApJ...349..126F, 1996ApJ...469..171G, 1998MNRAS.298...33R, 2006ApJS..165..283A, 2017MNRAS.466.4573S}. To date, the three-dimensional structure of \hii{} regions and surrounding medium has been studied mostly using two-dimensional maps of dust or molecular emission, using radial velocities of various atomic or molecular tracers \citep[e.~g.][and many others]{Emprechtinger_2009, Anderson_2015, Anderson2019, 2020A&A...639A...2P, 2022A&A...659A..77B}. In some cases for nearby objects, it is possible to use three-dimensional maps of dust extinction, complemented by 6D~data from GAIA, to reconstruct not only the structure, but also the star-formation history in molecular clouds \citep[e.g.][]{2023ApJ...947...66F}.

The present study continues the idea of using combinations of various dust tracers to reconstruct the three-dimensional structure of \hii{} regions, as in our previous work using archival data on far-infrared (IR) dust emission and optical wide-field images for the \hii{} region Sh2-235 \citep[][Paper~I hereafter]{Kirsanova_3D}. Here, we develop the ideas used in Paper~I for exploring the three-dimensional structure for more extensive work on several \hii{} regions in the star-forming complex S254-258 (see Appendix~\ref{sec:appendixA} for a review of the appearance and estimated physical parameters of the individual \hii{} regions). This complex appears as six bubbles in the mid-infrared (see Fig.~\ref{fig:large_scale}), with five of them associated with the optical \hii{} regions Sh2-254, Sh2-255, Sh2-256, Sh2-257 and Sh2-258 (S254, 255, 256, 257 and 258 hereafter), as well as a faint region in the north-east which has not catalogued yet. The complex is situated in Perseus spiral arm, which has a distance from the Sun of $\approx 2$~kpc in the direction of the anti-centre of the Galaxy \citep{2006Sci...311...54X, 2014ApJ...790...99C}. The the gas and dust temperatures in the S254-258 star-forming complex are mostly maintained by massive B-type stars \citep{Evans_1977, Sargent_1981, Avedisova_1984, Heyer_1989, Dors_2003}. As the majority of the massive stars which generate infrared `bubbles' in the Galaxy have spectral classes from O to B2/B3 \citep[e.~g.][]{Churchwell_2006, Deharveng_2010}, our complex can be considered as a typical example of such structures. The parameters of the \hii{} regions already measured in the literature are presented in Table~\ref{tab:HII_pars_lit}.

Locally, the energetics of the complex are supported by young embedded stellar objects surrounding the \hii{} regions, with the most prominent star-forming region located between S255 and S257, as well as to the south of them (see large-scale maps by~\citet{Chavarria_2008, Bieging_2009} and Fig.~\ref{fig:large_scale}). These authors suggest that a sequential star-formation process in the complex took place in the intersection of the bubbles associated with the \hii{} regions \citep[see also][]{Ojha_2011, Chavarria_2014, Kohno_2022}. In particular, star formation in the most massive and prominent region at mm and infrared wavelengths is a molecular ridge at the interface of S255 and S257 \citep[e.g.][]{2011A&A...527A..32W,2015ApJ...810...10Z, 2018ARep...62..326Z}, and was probably induced by the expansion of these \hii{} regions \citep{Bieging_2009, Mucciarelli_2011, Ojha_2011, Wang_2011, Zinchenko_2012, Ladeyschikov_2021}. \citet{Samal_2015} and \citet{Ryabukhina_2018} found a 20-pc long star-forming molecular filament to the south of the bright optical nebulae, where the compact \hii{} region S258 is embedded. Finally, the compact \hii{} region S256 is situated to the south-west of S257, on the edge of an extended molecular cloud.

\section{Observations and data reduction}\label{sec:obs}

We conducted optical observations with both the 6-m and 1-m telescopes of the Special Astrophysical Observatory of the Russian Academy of Sciences (SAO RAS), using a similar technique described in Paper~I. Here, we present a detailed description of the observations and data reduction steps, with special attention paid to differences in the detectors used and steps of the data reduction.

\subsection{MaNGaL emission line mapping}

Observations with the 1-m Zeiss-1000 telescope of SAO RAS were performed using the Mapper of Narrow Galaxy Lines \citep[MaNGaL,][]{mangal}, which is a tunable-filter photometer based on a piezoelectric scanning Fabry-Perot interferometer (FPI) working in a low order of interference ($\sim20$ at 656 nm). The width of the instrumental profile in the 480--700 nm spectral range  was FWHM=$12\pm1$\AA. The scanning system allows setting the central wavelength (CWL) of the   FPI transmission peak to the desired wavelength at the centre of the field of view (FOV). A bandpass filter (FWHM was 10~nm in the  \Hbeta{} and 25~nm in other spectral ranges) is used to select a single peak of the FPI transmission. Compared with the previous study of S235 in Paper~I, the current work had the following differences in observation technique. We used the Tektronix $1K\times1K$ CCD detector with 24~$\mu m$ pixel size, operated by the new DINACON-V controller developed at SAO RAS \citep{Ardilanov2020gbar.conf..115A, Afanasieva2023}, which provides a FOV of 11.5 arcmin (vignetted by a round holder of a bandpass filter) at a sampling of 0.9 arcsec~px$^{-1}$. This is significantly larger than the square FOV ($8.7\times8.7$ arcmin) in the observations of S235 in Paper~I with the previous detector.  In Paper~I we obtained relatively deep images in each emission line using the MaNGaL `single image mode', where the CWL was switched between the emission line wavelength and the neighbouring continuum. Here, the `scanning mode' was used to correct the mismatch between the peak CWL and the emission line barycentre caused both by variations of the object Dopler velocities and instrumental CWL distribution. We quickly scanned the wavelength regions around the emission line with a short exposure and lower spatial resolution with $4\times4$ readout binning.  However, the radial  change of the peak CWL across this large FOV with the TK1024 detector was quite significant (exceeding  1/2 of the FWHM). For this reason, all observations of S255-257 region were performed only in a scanning mode, with $1\times1$ binning and relatively long exposures. During a scanning cycle we obtained several (5--12) images with CWL steps of 5--7\AA{}, covering the spectral range around each emission line (\Hbeta, \OIII$\lambda5007$) or line system (\Halpha+\NII$\lambda6548,6583$, \SII$\lambda6717,6731$). The spectral continuum was sampled at 20--30\AA{} from each emission line region. The exposure time of each frame was 600~s for \Hbeta and 120--200~s for the other lines, and each cycle was repeated several times in order to average the contribution of atmospheric parameters and air mass variations. The total exposures, seeing value and number of frames in each scanning cycle ($n_z$) are given in Table~\ref{tab:Z1000_obs}. 

The data reduction was performed using the programs and algorithms described in Paper~I and \citet{mangal}. The calibration to an absolute intensity scale was performed using spectrophotometric standard stars observed in the same spectral range immediately before or after the exposures of the nebulae. The products of the data reduction are data cubes containing $n_z$-channel low-resolution spectra (FWHM=12\AA{}, or $R=\lambda$/$\delta\lambda=$400--650) at each pixel. The data cubes were aligned using the astrometric calibration from the astrometry.net software \citep{Lang2010AJ....139.1782L}, and the data cubes in the \Hbeta{} and \OIII{} emission lines obtained on different nights were co-added. 

The continuum-subtracted spectra in each data cube were fit with a Lorentzian profile, providing a good approximation of FPI instrumental profile. We used a one-component Lorentzian for the  \Hbeta{} and \OIII{} lines, double-component model for \SII{}, and three-component model for the \Halpha+\NII{} line systems. The free parameters were the central velocity, amplitude and FWHM (the same for all lines in the cases of \SII{} and \Halpha+\NII). For the line systems the wavelength difference between lines were fixed, and in the case of the \NII{} doublet the line ratio (1:3) was also fixed. The result of this procedure is two-dimensional maps of the flux in each emission line. The signal-to-noise (S/N) maps were calculated as a ratio of a line amplitude to photon noises in each pixel.  In order to improve the detection limit for the faint structures we also produced data cubes with different binning: $1\times1$, $2\times2$ , $4\times4$ and $8\times8$ pixels.

\begin{table}
	\centering
	\caption{MaNGaL observations at the Zeiss-1000 telescope}
	\label{tab:Z1000_obs}
	\begin{tabular}{llrrr} 
		\hline
		Date  & Sp. range& Exposure, s &  nz &seeing, $''$\\
		\hline
   	2020 Mar 01& \Hbeta{}  & 5400      &  5 &2.6 \\
    2020 Mar 02& \Hbeta{}  & 4800      &  5 &2.6 \\
    2020 Dec 24& \OIII{}  & 4120      &  7   &2.0 \\
    2020 Dec 24& \Halpha+\NII  & 3600& 10 &  1.9 \\
    2022 Nov 28& \SII{}  & 8000& 12 &  1.5 \\
    2022 Nov 30& \SII{}  & 15800& 12 &  1.2 \\
	\hline
   \end{tabular}
\end{table}

\begin{table}
	\centering
	\caption{SCORPIO-2 observations at the BTA-6 telescope}
	\label{tab:6m_obs}
	\begin{tabular}{llrr} 
		\hline
		Date  & region and PA, deg& Exposure, s &  seeing, $''$\\
		\hline
   	2023 Jan 16& S255, PA=109 & 1200     &2.3 \\
    2023 Jan 16& S255/257 blank & 600     &2.3 \\  
    2023 Jan 16& S256, PA=170 & 720     &1.9 \\
    2023 Jan 16& S256 blank & 600     &1.9 \\  
    2023 Jan 16& S257, PA=87 & 1200     &2.2 \\
    2023 Jan 16& S258, PA=87  & 1200     &1.9 \\  
	\hline
   \end{tabular}
\end{table}

\subsection{SCORPIO-2 spectroscopy}

For additional study of the ionized gas properties, and also to check the calibration of the MaNGaL flux maps, we performed observations in the long-slit mode of the SCORPIO-2 multi-mode focal reducer \citep{AfanasievMoiseev2011} on the prime focus of the 6-m Big Telescope Alt-Azimuth (BTA) telescope. The slit had a length of 6.3 arcmin and width of 1 arcsec, which provides a spectral resolution about 5\AA{} in the range 365--730nm. The E2V 261-84  $2K\times4K$ pixel CCD detector gave a spatial sampling 0.39 arcsec per pixel, the detailed description of this CCD camera is presented in \citet{Afanasieva2023}. The spectra were obtained along 4 positions in S255--258 regions, and the corresponding position angles and other parameters are listed in the Tab.~\ref{tab:6m_obs}. The preliminary data reduction was performed in a standard way using custom developed IDL-based software, as described in our previous works \citep[e.g.][]{Egorov2018MNRAS.478.3386E}. To remove the contribution from airglow lines to the observed spectra where the ionized gas emission was detected along the entire slit, two blank fields were exposed at a distance of $\sim1$ degree from the S254-258  complex. In the case of the compact S258 region, the sky night spectrum was taken from the area free from object emission along the slit. To calibrate the reduced spectra to an absolute flux density scale, we used the spectra of spectrophotometric standard star observed at a close zenith distance immediately after the observations of the nebular complex. The integrated fluxes of emission lines were obtained via single-component Gaussian fitting after removal of continuum emission fit by splines. The uncertainties of the measured fluxes were estimated from Monte Carlo simulations of the synthetic spectra with the given S/N.

\section{Methods}
\label{sec:methods}

\subsection{Properties of ionized gas}

\subsubsection{Electron temperature}

Unfortunately, the commonly used temperature-sensitive line ratio \OIII~$\lambda4363$/($\lambda4959$+$\lambda5007$) could not be applied, as the weaker \OIII~$\lambda$4363 emission line was not detected in the integrated spectra of the \hii{} regions in the S254--258 star-forming complex. Instead, we estimated $T_e$(\NII) using the ratio of nebular to auroral nitrogen line intensities $Q_{2,N}=$ \NII ($\lambda6548$ + $\lambda6584$)/$\lambda5755$ and the improved calibration relation between $T_e$(\NII) and $Q_{2,N}$ for the low density regime (\nelec{}$<100\,\mbox{cm}^{-3}$) according \citet{Pilyugin2010ApJ...720.1738P}. Because the $Q_{2,N}$ ratio depends weakly on the interstellar extinction \AV, we dereddened the line intensities in the integrated spectra using the H$\alpha$/H$\beta$ line ratio, as described in Sec.~\ref{sec:extdens}. The ratio of the Balmer lines also depends on $T_e$, so we did several iterations to determine the $T_e$(\NII) values for S255, S256, and S257. The integration radii were 7--120~arcsec in S255 and S257, and 7--50~arcsec in S256. The zone of $r<7$ arcsec around the bright central star was ignored to avoid contamination.

\subsubsection{Extinction and electron density}
\label{sec:extdens}

In our analysis we used only those pixels of the maps and velocity channels in the spectra where the S/N level was ~$>3$ and the line intensities and their ratios are not smaller than their uncertainties. Our methods are the same as in Paper~I, therefore, here we only briefly summarise them. The basic property of the \hii{} regions, electron density \nelec, was calculated using the ratio of the [\ion{S}{II}]~$\lambda\lambda$6716\AA{}/6731\AA{} lines and Eqs.~(3) and (4) of \citet{Proxauf_2014}. We found \nelec{} in the entirety of the S255, 256, S257 and S258 \hii{} regions, and partially in S254 in the north-west part of the MaNGaL images. No spectroscopy was performed for S254 due to the weak emission. We determined the interstellar extinction \AV{} comparing the observed H$\alpha$/H$\beta$ intensity ratio for the each pixel of the MaNGaL images or position on the slits with the intrinsic ratio for Case~B conditions given by \citet{Osterbrock06} and the reddening law of \citet{Cardelli89}.

For the nebular spectra, we determined the extinction also using H$\gamma$/H$\beta$, H$\delta$/H$\beta$ and H$\epsilon$/H$\beta$ values. For the ratio of total to selective extinction, we adopted the value of $R_V=3.1$ as our standard model. However, we tried higher values up to $R_V=5.5$ for the analysis of the spectra. In order to estimate uncertainties of \nelec{} and \AV, we applied a bootstrap approach using the measured flux densities and their uncertainties for a random sample. In order to study gradients of these values over the nebulae and increase the values-to-uncertainties ratios, we binned the original images and finally used an 8 times larger pixel for the maps of physical parameters and dereddened images of the surface brightness. For the spectra, we used 20 times larger bins to improve the signal-to-noise ratio.

\subsection{Dust column density}

In order to study the three-dimensional structure of the \hii{} regions, we used the map of dimensionless equivalent of the optical extinction $A_V^{\rm IR}$ obtained by \citet{Ladeyschikov_2021}. This map can be transferred into hydrogen column density as $A_V^{\rm IR} = 5.3 \times 10^{-22} \times$\,$N$(\ion{H}{I}+ H$_2$)~mag \citep{Bohlin_1978, Rachford_2009}. The map we use in the present study is based on a grey-body fit of the spectral energy distribution (SED), with a dust emissivity $\kappa \propto \lambda^{-1.8}$, of the {\it Herschel} Hi-GAL data \citep{Molinari_2010} in the 160-500~$\mu$m range.

\subsection{Properties of ionizing stars and their extinction}

We  used the stellar spectra in order to estimate the spectral type of the ionizing stars in each of the \hii{} regions. Furthermore, the shape of the SED of the ionizing stars at optical and near-infrared wavelengths can be used to constrain the value of $R_V$, i.e. the ratio of the total to selective extinction ($A_V$/$E_{B-V}$), in the extinction law, and provide an independent measure of the extinction $A_V$ towards the star.

We fit the continuum-normalized spectra of the ionizing stars of S255, S256 and S257 using the `Tlusty' grids of non-LTE stellar atmospheres for O- \citep{Lanz_2003} and B-stars \citep{Lanz_2007}, convolved to the spectral resolution of our observations (FWHM=5\AA).  As a first step, we initially placed no restrictions on metallicity $Z/Z_\odot$, surface gravity $\log{g}$ or effective temperature $T_\mathrm{eff}$ of the model spectra, and found the best fit (i.e., lowest $\chi^2$) from the OSTAR2002 or BSTAR2006 precomputed grids without performing any interpolation on the stellar parameters.  We specifically excluded wavelengths with nebular emission lines (which might leave artefacts after the background subtraction), DIBs and atmospheric absorption from the fits. For all three regions, we found that a metallicity of $Z$/$Z_\odot=0.2$ and surface gravity of $\log{g=4}$ (typical for main-sequence OB stars) provide the best fits, and we therefore fixed these values for the rest of the fitting process.

After fixing the metallicity and surface gravity, we fit the effective temperature in a more refined manner using the equivalent widths of \ion{He}{I} and \ion{He}{II} lines, as well as the \ion{C}{III} 4650\AA{} line.  This approach has the advantage that it is less sensitive to the exact shape of the stellar continuum used for normalization. We interpolated the Tlusty continuum-normalized spectra to arbitrary effective temperatures, as the two grids have spacings of either 2500~K (the OSTAR2002 grid) or 1000~K (the BSTAR2006 grid). The uncertainty in the temperature determination was estimated by a bootstrapping process, where we created synthetic spectra from the observed spectra with flux values drawn from a normal distribution with a width equal to the estimated flux uncertainty of the observations (typically about 10\%).

In order to determine the extinction parameters $A_V$ and $R_V$ from the SED of the ionizing stars, we compared the best fit from the Tlusty spectra together with photometric measurements in the optical and near-infrared. Specifically, we used $BV$ measurements from UCAC5 \citep{Zacharias_2017}, $G$, $G_\mathrm{BP}$ and $G_\mathrm{RP}$ from Gaia DR2 \citep{Gaia_DR2}, $gri$ from APASS \citep{APASS}, and $JHK$ from 2MASS \citep{Skrutskie_2006}.  The observed (reddened) SED was compared with the theoretical SED using the reddening law of \citet{Cardelli89}, with $A_V$ and $R_V$ as free parameters, and their uncertainties were found using a bootstrapping process. This procedure has the advantage of providing an additional, independent check on the values of $A_V$ derived from the H$\alpha$ and H$\beta$ images of the nebulae from MaNGaL, and allows us to determine $R_V$, which cannot be determined independently from $A_V$ from observations of only two Balmer lines.

\section{Results of observations and analysis}

The resulting integrated spectra reveal numerous ionized gas emission lines: the hydrogen Balmer series (\Halpha, \Hbeta, H$\gamma$, H$\delta$, H$\varepsilon$), as well as forbidden lines (mostly in S255 and 257): \OI$\lambda6300$, \OII$\lambda3727$,  \OIII$\lambda4959,5007$,   \NII$\lambda5199,5755,6548,6583$, \SII$\lambda6716,6731$, \ArIII$\lambda7136$. Also in S255 and 257 the helium lines \HeI$\lambda5876,6678$  were detected. The integrated spectra and relative flux of all emission lines are presented in Appendix~\ref{sec:appendixB}.

Electron temperatures $T_e$, obtained from the integrated spectra, are presented in Table~\ref{tab:electrontemperatures}. These values are in good agreement with results measured in the literature earlier (Table~\ref{tab:HII_pars_lit}). Also in Table~\ref{tab:linefluxes_dered} we present $T_e$(\NII) values  calculated separately for the inner and outer parts of the S255 and S257 regions. A possible radial trend of $T_e$ is not seen. We used a value of $T_e=7940$~K, averaged over all the \hii{} regions, to create maps of the ionized gas parameters based on the MaNGaL images.

\begin{table}
    \centering
    \begin{tabular}{ccccc}
    \hline
       \hii{} region  & $T_e$ & $T_\mathrm{eff}$ & $A_V$ & $R_V$ \\
                      & (K) & (K) & (mag) &  \\
    \hline
        S255 & $8141\pm90$ & $31300\pm120$ & $3.84\pm0.12$ & $3.10\pm0.32$ \\
        S256 & $8222\pm762$ & $22750\pm270$ & $5.61\pm0.15$ & $4.66\pm1.33$ \\
        S257 & $7459\pm211$ & $27550\pm130$ & $2.76\pm0.13$ & $3.30\pm0.45$ \\
       \hline
    \end{tabular}
    \caption{Electron temperatures obtained from the integrated spectra ($T_e$), effective temperature of the ionizing stars ($T_\mathrm{eff}$) and parameters of the absorbing medium toward the stars derived from the observed spectra and SEDs ($A_V$, $R_V$).}
    \label{tab:electrontemperatures}
\end{table}

\begin{figure}
	\includegraphics[width=\columnwidth]{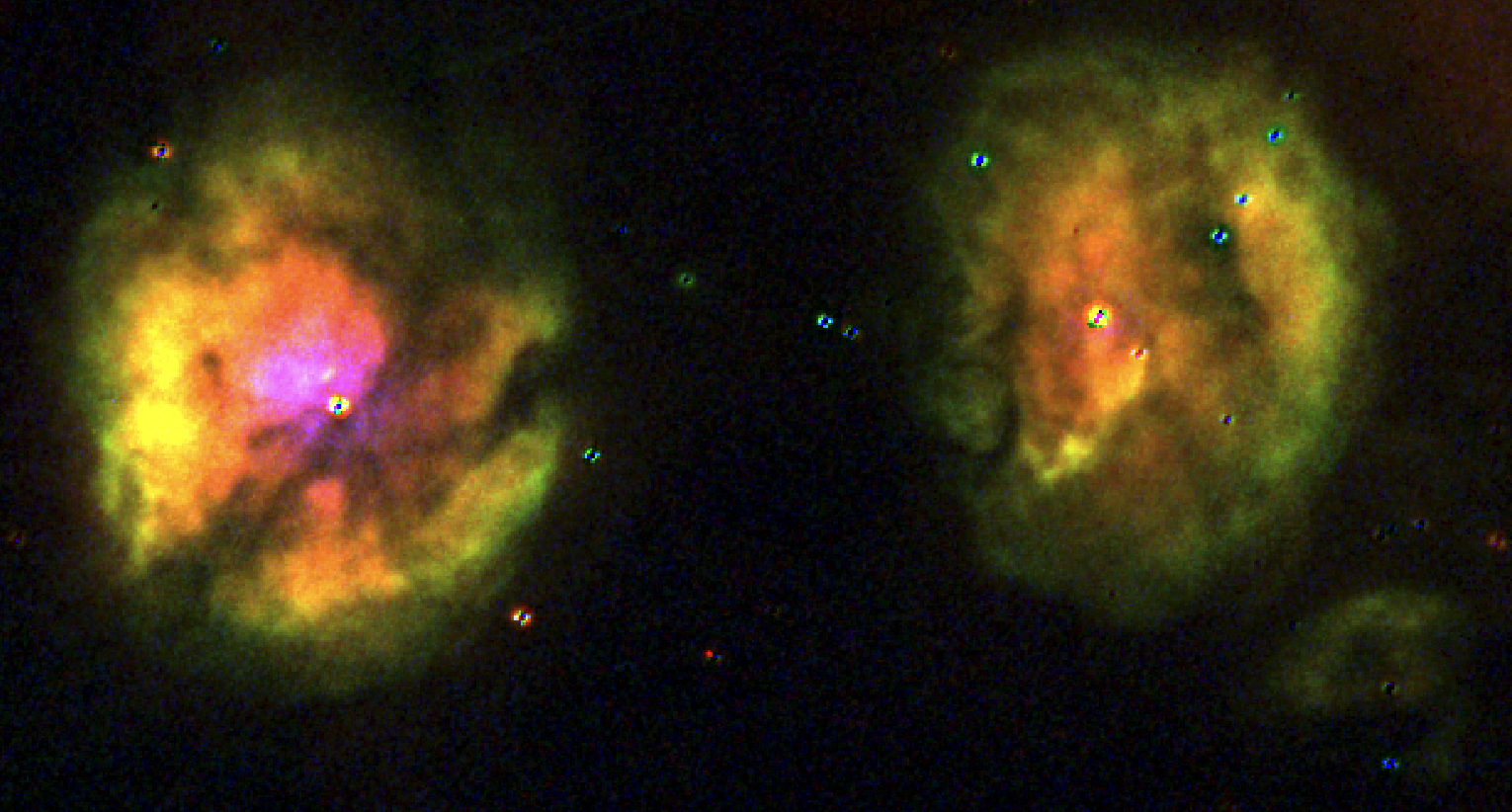}\\
    \caption{Colour composite image using the surface brightness of three emission lines, with red: \Hbeta, green: \SII, and blue: \OIII.} 
    \label{fig:threelines}
\end{figure}

\begin{figure}
	\includegraphics[width=\columnwidth]{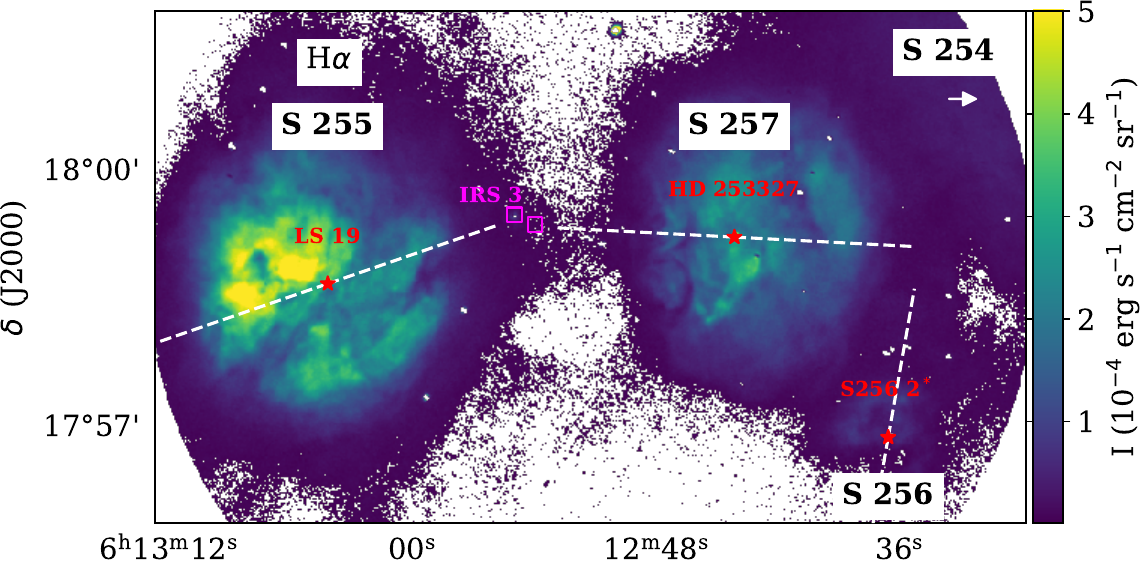}\\
	\includegraphics[width=1.03\columnwidth]{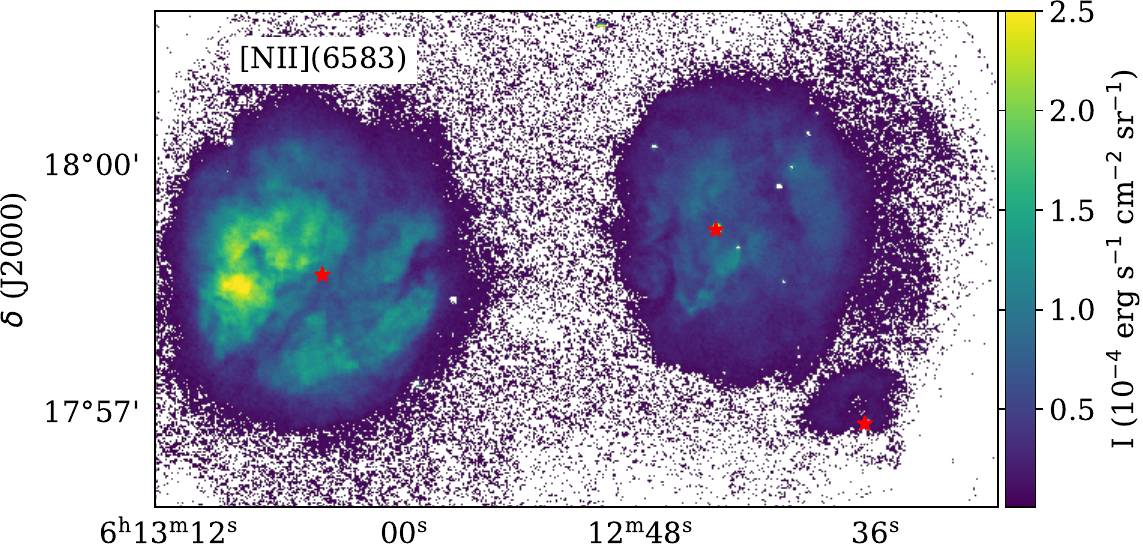}\\
	\includegraphics[width=1.03\columnwidth]{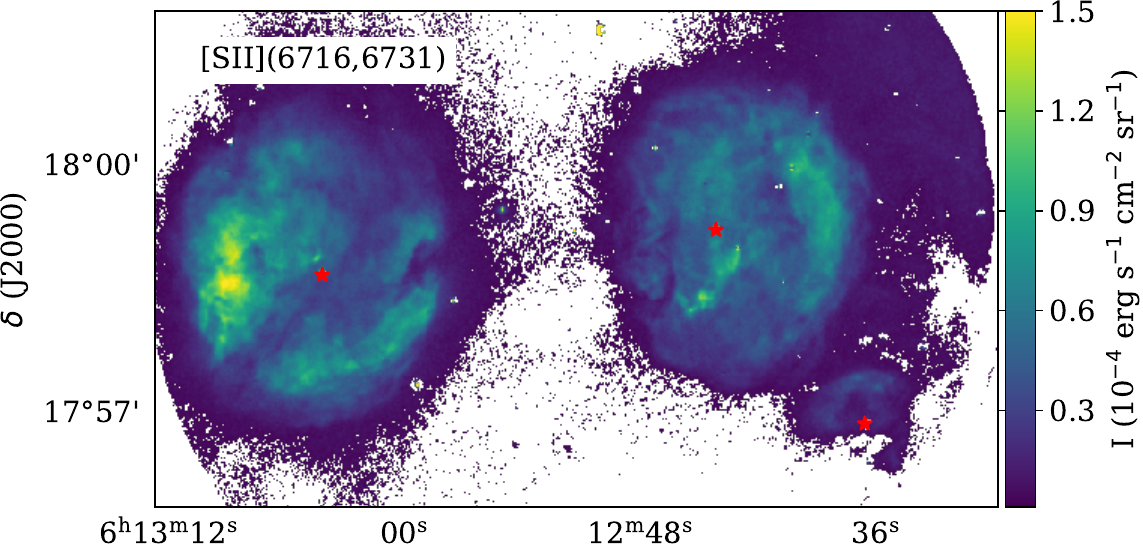}\\
	\includegraphics[width=1.03\columnwidth]{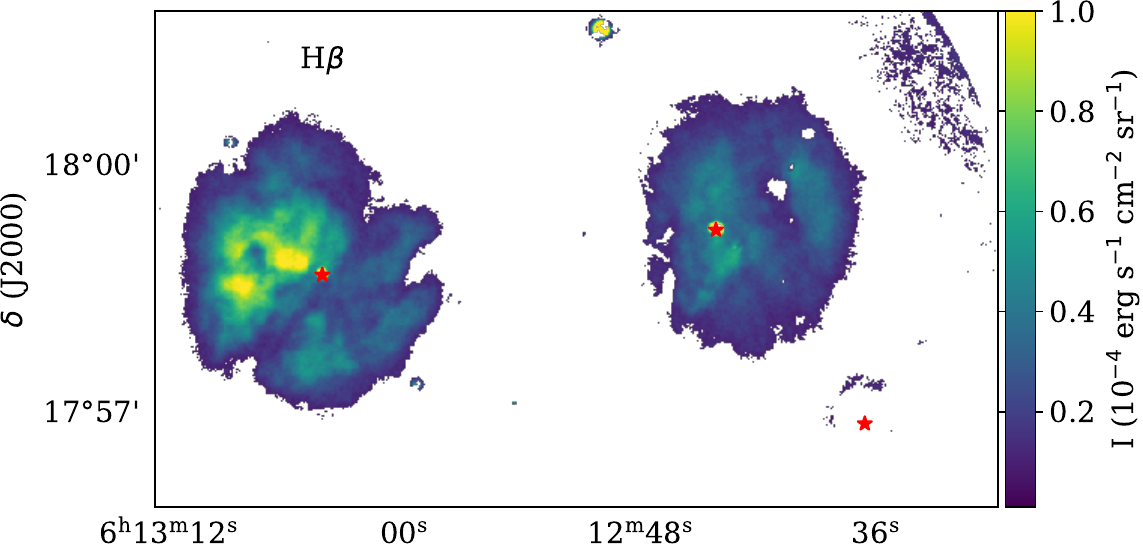}\\
	\includegraphics[width=\columnwidth]{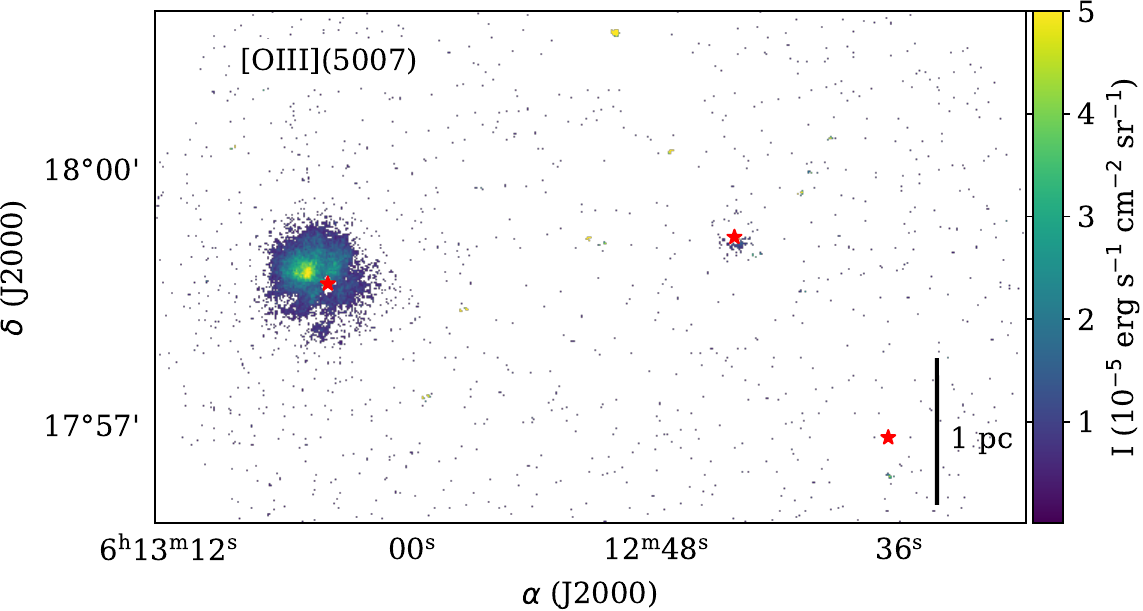}
    \caption{MaNGaL images of S254-258 star-forming region. Only pixels with S/N ratio~$> 3$ are shown. The red star symbols show the positions of the ionizing stars. The names of the \hii{} regions and their ionizing stars are shown by black and red, respectively, on the top panel. The IR sources S255~IRS1 and IRS3 are shown in magenta, where we used the coordinates from \citet{1997ApJ...481..327H}. The white dashed lines show positions of the slits for spectra. A physical scale bar is shown by black line on the bottom panel.} 
    \label{fig:obsres}
\end{figure}

A colour composite image using the surface brightness of three emission lines without continuum is shown in Fig.~\ref{fig:threelines}. Separate images of the surface brightness of the hydrogen recombination lines \Halpha{}, \Hbeta{}, and forbidden lines of metals \SII, \NII{} and \OIII{} are shown in Fig.~\ref{fig:obsres}. The images are shown scaled to the values of their surface brightness, where the \Halpha{} lines reaches $5\times10^{-4}$~erg~s$^{-1}$~cm$^{-2}$~sr$^{-1}$ and the \OIII{} line reaches a factor of ten less in the brightest parts. Besides the three \hii{} regions S255, S256 and S257 situated completely within our field of view, we also detect diffuse emission from the east edge of the \hii{} region S254. The spatial distribution of the surface brightness is inhomogeneous over the images, with the brightest values to the north-east of the ionizing star LS\,19 in S255. There are several elongated structures, indicating that the absorbing material is concentrated in the central part of the image. Images of S256 in the \Halpha, \NII{} and \SII{} lines appear as rings around the ionizing star. The surface brightness of the \SII{} line in S255 also has a semi-ring distribution. The \OIII{} line is detected only to the north-east of LS\,19 in S255 and appears as a compact bright spot.

\begin{figure}
    \includegraphics[width=\columnwidth]{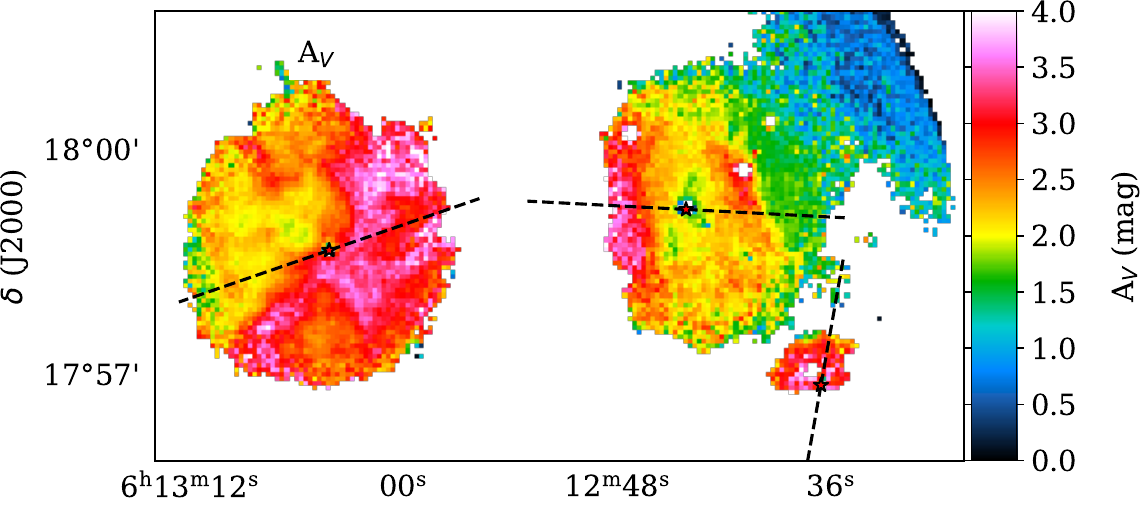}
    \includegraphics[width=1.015\columnwidth]{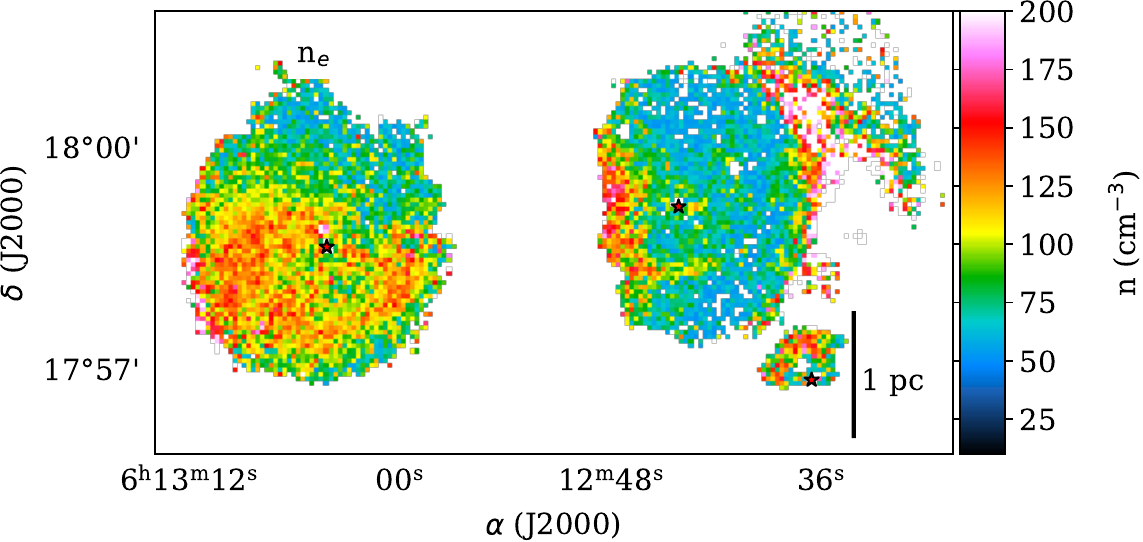}
    \caption{Physical parameters in S255-258: \AV{} (top) and \nelec{} (bottom). The images were rebinned to a pixel size 8~times larger than the original pixel in Fig.~\ref{fig:obsres}. The red star with black border symbols show the positions of the ionizing stars. The black dashed lines show the positions of the slits for the spectra. Pixels with values determined for \nelec{} but without \AV{} value were removed from the bottom map to simplify comparison. A physical scale bar is shown by black line on the bottom panel.
    }
    \label{fig:physparam}
\end{figure}

\begin{figure*}
    \centering
    \includegraphics[width=\columnwidth]{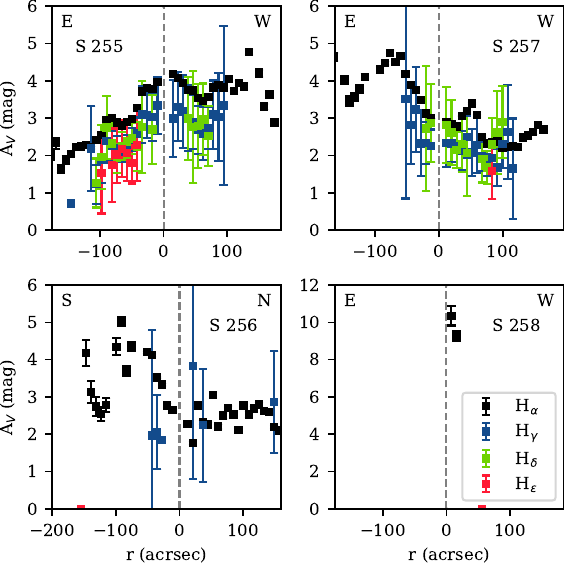}
    \includegraphics[width=\columnwidth]{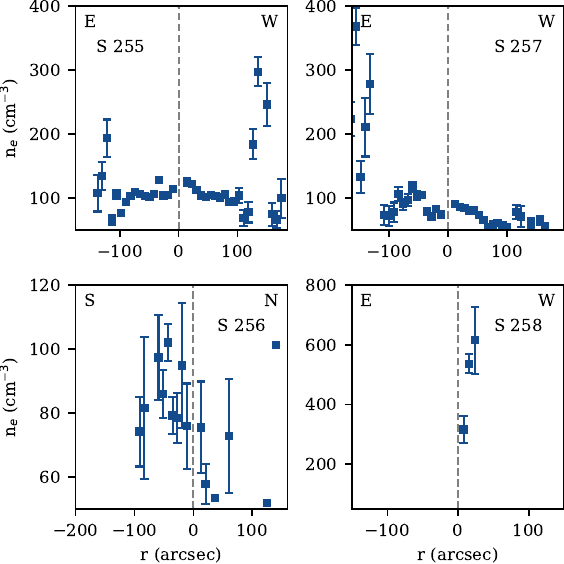}
    \caption{Extinction (left) and electron density (right) in the \hii{} regions along the slits. Vertical dashed lines show the location of the ionizing stars. The orientation of the slits in Fig.~\ref{fig:obsres} is shown by 'N', 'S', 'E' or 'W' symbols. Extinction values obtained with different Balmer lines are shown by colours.}
    \label{fig:results_from_spectra}
\end{figure*}

The spatial distribution of \AV{} is shown in Fig.~\ref{fig:physparam}. The optical extinction reaches 4~mag in the dark dust fibres in front of the south-west part of S255. The north-east part of S255 is less obscured, with $A_V \approx 1.5-2$~mag. Absorbing material is concentrated to the east in S257. The \AV{} value decreases from 4 to 1~mag to the west and north-west of the nebula. S256 is obscured up to $A_V=4$~mag, with the extinction peak located in the direction of the proposed ionizing star. The spatial distribution of the extinction with the enhancement between S255 and S257 is also visible in Fig.~\ref{fig:results_from_spectra}, where we show results from the analysis of the long-slit spectra; the slit spectra allow determining the \AV{} values even in positions where the signal on the MaNGaL images was too weak. \AV{} reaches 5~mag on the edge of the molecular cloud between S255 and S257, and has a similar value to the south of S256. S258 is deeply embedded into neutral absorbing material, with $A_V=9-11$~mag, but the optical emission around the ionizing star is weak and we did not obtain physical parameters in the rest of positions along that slit.

The extinction, obtained using other lines of the Balmer series: H$\gamma$, H$\delta$ and H$\epsilon$, is also shown in Fig.~\ref{fig:results_from_spectra}. Using the same reddening law, we find the \AV{} values systematically lower than the extinction from the ratio of H$\alpha$/H$\beta$ in S255 and S257, while high uncertainties do not allow concluding about S256. Using another total-to-selective extinction ratio $R_V$ value, we were able to bring the \AV{} values from different lines into agreement at $R_V = 4$. However, we do not exclude that the described effect is related with deviations of the observed line fluxes from the intrinsic values given by theory due to e.~g. inhomogeneity of the absorbing material. Our suggestion that the $R_V$ value can be higher than commonly used for diffuse medium is marginally supported by analysis of the spectra of the ionizing stars. The upper limits of $R_V$ toward them agree with the Balmer series analysis (Table~\ref{tab:electrontemperatures}). As the present study will be followed by a more extensive work on numerous \hii{} regions with various properties, we plan to provide a more general analysis of the $R_V$ value toward \hii{} regions in the following study.

The distribution of the electron density in the observed \hii{} regions is not homogeneous. The ionizing stars of S255 and S257 are surrounded by gas with $n_e \approx 100-150$~cm$^{-3}$ and $n_e \approx 50-75$~cm$^{-3}$, respectively (Fig.~\ref{fig:physparam}). We find that the southern part of S255 has systematically higher gas density compared to the northern part, where $n_e$ drops to $ \approx 75-90$~cm$^{-3}$. In S257, we find an east-west gradient of $n_e$ from $>100$~cm$^{-3}$ to $<70$~cm$^{-3}$. The largest \nelec{} value up to $400$~cm$^{-3}$ is found on the borders of the \hii{ } regions (Fig.~\ref{fig:results_from_spectra}). 

As is known from previous studies, there is a molecular cloud between S255 and S257. The central enhancement of the \nelec{} close to the molecular cloud is accompanied by a rise of the \AV{} value (see results from spectra in Fig.~\ref{fig:results_from_spectra}). This effect is less visible in Fig.~\ref{fig:physparam}, as we removed all pixels where $n_e$ is known but \AV{} is not, in order to make these maps more convenient for comparison. Therefore, these \hii{} regions irradiate the cloud from opposite sides and ionize the dense material. The enhancement of \nelec{} up to 200~cm$^{-3}$ in the outskirts of the \hii{} regions can also be related to the ionization of the surrounding dense neutral envelopes (see Fig.~\ref{fig:large_scale}). Therefore, the \hii{} regions S255 and S257 are reliable examples of media where UV photons penetrate through the dense medium. The widths of the dense ionized walls of the \hii{} regions are 20-30~arcsec, which is equivalent to $0.2$~pc for the distance of the S254-258 complex.

In the compact \hii{} region S256 we found $n_e \sim 100$~cm$^{-3}$. It seems to still be embedded in the parental molecular cloud. There is a north-east -- south-west gradient of the $n_e$ and \AV{} values from 100 to 50~cm$^{-3}$ (the limit for the diagnostic using the \SII{} lines) and from 4-5 to 2~mag, respectively. Another compact \hii{} region S258 demonstrates the highest $n_e$ up to 600~cm$^{-3}$, and \AV{} up to 9-11~mag. The optical spectroscopy allowed us to estimate the physical conditions only toward the ionizing star due to faintness of the lines.

\begin{figure}
        \includegraphics[width=\columnwidth]{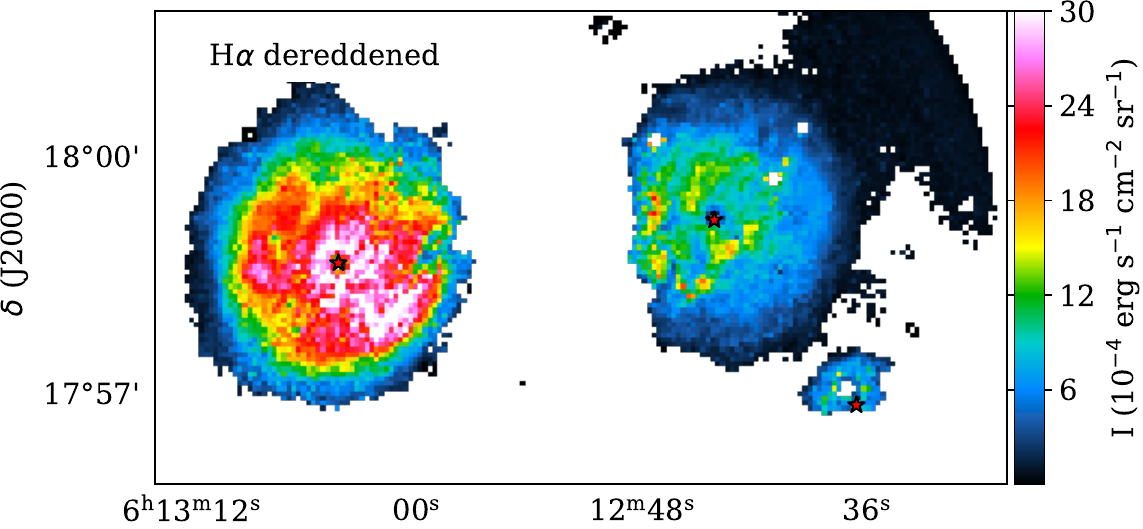}
	\includegraphics[width=\columnwidth]{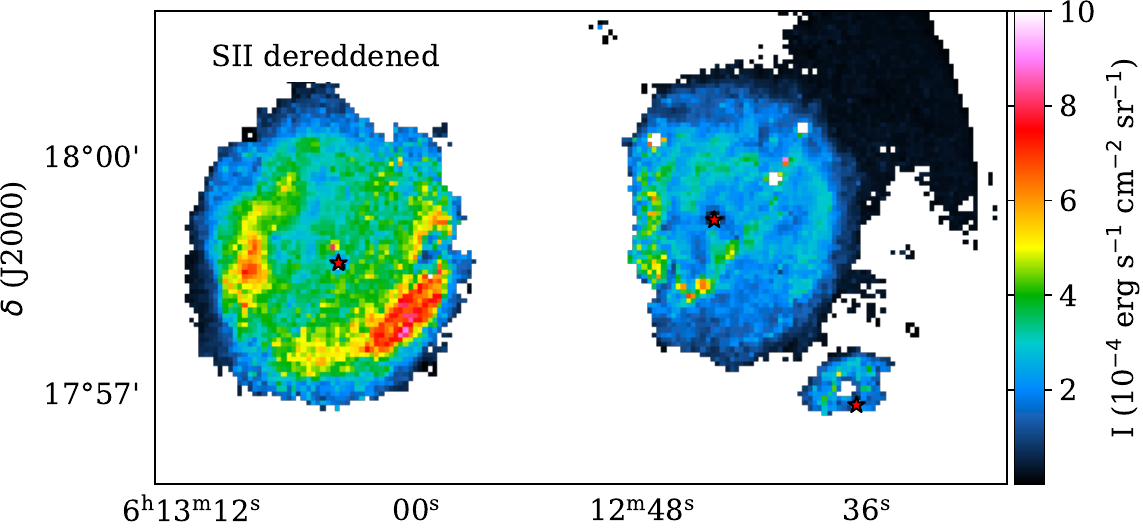}
	\includegraphics[width=\columnwidth]{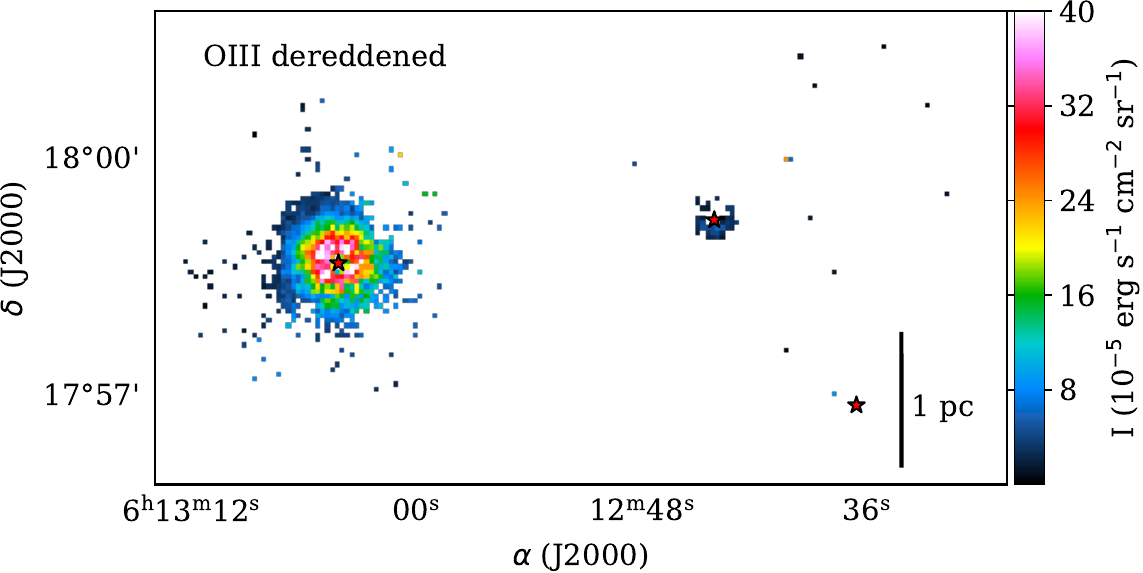}
    \caption{Dereddened images of the nebulae. Red star with black border symbols show the positions of the ionizing stars. The images are rebinned to a pixel size 8~times larger than the original pixel. A physical scale bar is shown by black line on the bottom panel.}
    \label{fig:dered}
\end{figure}

Applying the \AV{} map to the images of the line emission, we obtain the dereddened flux $F\left(\lambda\right)=F_\mathrm{obs}\left(\lambda\right) \times 10^{0.4 A_\lambda}$ in each filter, where the functional form of $A_\lambda$/$A_V$ was taken from the extinction law of \citet{Cardelli89}, and we used the \AV{} value derived at each pixel from the H$\alpha$/H$\beta$ line ratio.  The maps of dereddened line emission are shown in Fig.~\ref{fig:dered}. We find three different types of the surface brightness distribution among them. The dereddened \Halpha{} emission is distributed more uniformly in S255 than on the uncorrected maps (Fig.~\ref{fig:obsres}); now it has a peak of $3\times10^{-3}$~erg~s$^{-1}$~cm$^{-2}$~sr$^{-1}$, corresponding to the position of LS~19 and the bright area at the south-west. The dereddened \Hbeta{} and \NII{} images have qualitatively similar intensity distributions. The \SII{} line emission is distributed in a semi-ring way, also with a bright region of $8\times10^{-4}$~erg~s$^{-1}$~cm$^{-2}$~sr$^{-1}$ at the south-west. Due to re-binning of the original line emission images, the signal-to-noise level is higher. Therefore, we see larger area of the \OIII{} emission than in Fig.~\ref{fig:obsres}, and find the peak of the \OIII{} emission of $4\times10^{-4}$~erg~s$^{-1}$~cm$^{-2}$~sr$^{-1}$ towards the ionizing star. In S257, the \Halpha{} and \SII{} are distributed more uniformly at the level of 7 and  $2\times10^{-4}$~erg~s$^{-1}$~cm$^{-2}$~sr$^{-1}$, respectively, with moderate enhancements around the ionizing star and on the eastern border. 

\section{3D structure of the \hii{} regions}\label{sec:resobs}

Our conclusions about the clumpy and non-uniform medium are supported by early works in the infrared. \citet{Evans_1977} found that the spatial distribution of the neutral material around the \hii{} regions is inhomogeneous. Radio-continuum images by \citet{Israel_1976, Snell_1986, Fich_1993} show a centrally-peaked distribution of the radio continuum emission for S255 and S256, but diffuse emission for S254 and a peak shifted to the east for S257. They also found that the radio brightness in S255 is higher than in S257 by a factor of 2-3. Here we confirm a similar effect in the dereddened optical maps.

As shown above, distribution of the foreground material in S255 and S257 is inhomogeneous, with elongated structures in S255. These structures are visible also on the near-infrared images of S\,255 obtained by \citet{Kirsanova_KGO} but are not visible on the images in longer wavelength, e.~g. in mid-infrared or millimetre waves by \citet{Chavarria_2008, Bieging_2009}, respectively. A gradual decrease of extinction is observed in S257. Only in S256 we found the peak of \AV{} coincident with the location of the ionizing star.

\begin{figure}
\includegraphics[width=\columnwidth]{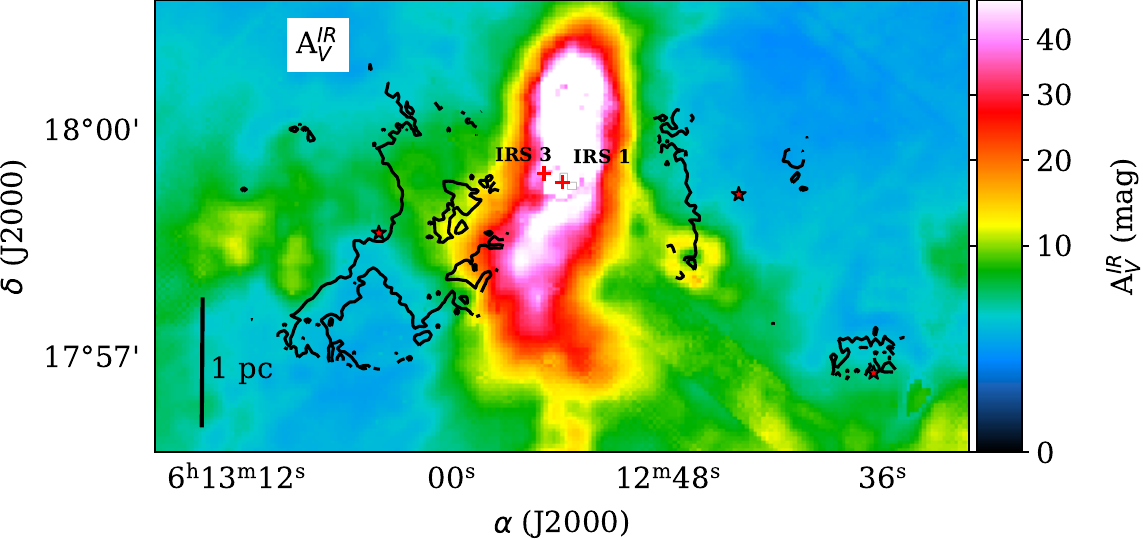}\\
\caption{Column density of neutral hydrogen in dimensionless equivalent of extinction values  $A_V^{\rm IR}$. Colour wedge has a square-root stretch to emphasise lower intensity features. Red star with black border show the positions of the ionizing stars. Black contours show the region with foreground $A_V \geq 3$~mag. The plus symbols show positions if infrared sources IRS~1 and IRS~3. A physical scale bar is shown by black line.}
\label{fig:herschel}
\end{figure}

In order to study the 3D structure of the \hii{} regions, we plot the neutral hydrogen column density in Fig.~\ref{fig:herschel} in dimensionless $A_V^{\rm IR}$ values. This procedure allows separating the neutral material of the front and the rear walls of the \hii{} regions. The difference between the \AV{} and $A_V^{\rm IR}$ values is that the first defines the the foreground extinction between the \hii{} region and the observer, but the second depends on three components: the foreground extinction, absorbing dust inside of \hii{} regions, and the neutral material behind the ionized region.

The most prominent feature in the distribution of the $A_V^{\rm IR}$ value corresponds to the molecular cloud between S255 and S257, where $A_V^{\rm IR} \geq 50$~mag and where active star formation is taking place (see relevant references in Sec.~\ref{sec:intro}). No optical emission is found towards regions where $A_V^{\rm IR} \geq 20$~mag, and our contours of $A_V \geq 3$~mag avoid that area. Comparing the $A_V^{\rm IR}$ and $A_V$ values in S255, we find inhomogeneities both in the front and rear neutral walls. Namely, there is an area to the south of the ionizing star where $A_V^{\rm IR} \approx A_V$, and we suggest that the rear wall in that direction is absent. The rear wall of this \hii{} region is visible as a continuous curved line from the west, north and east of LS~19 star. Thus, S255 appears as an example of an \hii{} region surrounded by neutral walls from all directions, which makes this object a promising target for comparison with theoretical models of \hii{} regions. 

Both S255 and S257 appears similarly, namely, they look as round \hii{} regions at optical wavelengths. The ionized gas is surrounded by neutral envelopes visible through the mid-infrared emission of PAHs particles \citep[see studies by e.~g.][about the appearance of bright PAH emission in the neutral envelopes of \hii{} regions, visible e.~g. by {\it Spitzer}]{2013ARep...57..573P, 2016ApJ...830..118S, 2016ApJ...819...65S, 2016A&A...586A.114M, 2018ApJ...858...67B, 2022MNRAS.509..800M, 2023IAUS..362..268K}. However, the morphology of S257 is different from those found in S255. There are no pronounced front or back walls in S257; this region is situated on the edge of the dense molecular cloud. We see neutral borders between S257 and S255, as well as between S257 and S254, in the {\it Spitzer} images, which implies that S257 is still bounded by neutral gas from at least the east and west sides. 
These observations are consistent with previous results by \citet{Bieging_2009}, who observed CO lines in the star-forming complex and found a good correlation between CO and mid-IR emission of warm dust. These authors suggested that S255 is probably on the near side of the CO molecular ridge, while S257 is carving out part of the molecular cloud. 

Relating the measured surface brightness to the line-of-sight depth using the emission measure, as in Paper~I, we find the depth of S257 is two times larger than in S255. The values of the depth are 10-20~pc, while the radii are $\approx 1$~pc. In actuality, the ionized volume of S257 can be smaller, due to escape of the gas from S257, as it is located on the border of the dense cloud. S255, on the other hand, is surrounded by dense neutral walls from all sides. We propose that the depth values can be overestimated with this simplified approach, probably due to the inhomogeneous distribution of the ionized gas along the line of sight. Alternatively, as the front and rear walls are non-uniform (S255) or almost absent (S257), the ionized gas can escape from the 1~pc vicinity of the ionizing stars, where 1~pc is the radius of the \hii{} regions in the plane of the sky \citep[see the studies already mentioned in Sec.~\ref{sec:intro} and also recent work by][]{Kirsanova_KGO}. 

Comparing the parallax-based distances of LS\,19 \citep[2060$^{2181}_{1951}$~pc by][]{Gaia_DR3} to the water maser in S255~IRS1 \citep[1780$^{1900}_{1670}$~pc by][]{2016MNRAS.460..283B} or to the methanol maser in the same direction \citep[1590$^{1660}_{1530}$~pc by][]{2010A&A...511A...2R}, we find that the ionizing star of S255 is at least 50~pc farther from the observer than the masers (compare the low limit 1951~pc for LS\,19 and the high limit for the water maser of 1900~pc). Therefore, the star-forming complex appears to be extended along the line of sight. For HD\,25332 the Gaia~DR3 distance has a quality much less than for LS\,19 and we consider a spectro-photometric measurement $2.46\pm0.16$~kpc by \citet{Russeil_2007} as the best measurement for now. This distance value supports our suggestion that S254-258 is extended perpendicular to the plane of the sky.

In S256, there is enhancement of the $A_V^{\rm IR}$ in the south in the direction of the molecular cloud, which is accompanied by a rise of the \AV{} and \nelec{} values. This \hii{} region, as well as S257, is apparently an example of a blister-type \hii{} region, which was formed by a star on the border of the dense molecular cloud. In these types of \hii{} regions, density of all the neutral as well as ionized gas components decreases away from the molecular cloud, as it was shown by simulations and observations e.~g. by \citet{1978A&A....70..769I, 1979MNRAS.186...59W, Tenorio-Tagle1979, ODell_2001, 2007A&A...464..995P, ODell_2009, 2012ApJ...745..158G} and many others for the Orion Nebulae and another \hii{} regions. We note that subsequent analysis of the ionized and neutral gas kinematics can be useful to confirm this suggestion, which will be done in forthcoming studies.

Having only two measurements of the \AV{} value in S258, we compared them with $A_V^{\rm IR}$ (not shown on the figures) and found $A_V^{\rm IR} < A_V$. Spatial resolution, which is more than two times higher in the optical data vs the infrared, possibly explains this finding. Fig.~\ref{fig:large_scale} shows that the mid-IR image of S258 has only a compact and bright spot near the ionizing star. The 3D structure of this \hii{} region can not be studied with the available data for now.

The spatial distribution of the dereddened \SII{} emission and \nelec{} demonstrate that S255 is not a pure `classical' spherical \hii{} region, where electrons are distributed uniformly, but resembles a ring-like structure, at least in the plane of the sky, as is sometimes observed for planetary nebulae \citep[e.~g.][]{Turatto_2002, ODell_2007}. The presence of the front and back neutral walls shows that the \hii{} region has no cavity free of electrons along the line of sight, and does not resemble a torus. This object resembles an ionized ellipsoid with non-uniform density distribution.

We applied a non-stationary model of an \hii{} region surrounded by a PDR \citep{Kirsanova_2009, Kirsanova_2019_rcw120} to the ionizing stars of S255 and S257 to simulate them and confine their physical conditions. Simulations with uniform, as well as with non-uniform, distribution of $n_{\rm init}$, where the density rises with distance, always produces uniform \hii{} regions. Therefore, we again suggest that the high electron density on the borders of \hii{} regions can be related to the penetration of diffuse UV~photons through clumpy neutral envelopes. Similar conclusions about clumpy PDRs were made e.~g.~by~\citet{Hogerheijde_1995, Lis_2003, Andree-Labsch_2017} for the nearby Orion~Bar PDR, and by \citet{Ciurlo_2019, Kirsanova_2020, Schneider_2021} for more distant regions, where the clumps were not resolved in the observations. 

Another possibility to obtain shell-like ionized volumes is to include stellar wind in our model. Many authors demonstrated how stellar wind changes the structure of uniform \hii{} regions and produce central cavities, e.~g.~\citet{2016A&A...586A.114M, 2019MNRAS.486.4947K, 2022MNRAS.517.4940W, 2023IAUS..362..268K}. We do not see such signatures of the wind as broad line wings in the stellar spectra. We can exclude wind only inspecting stellar spectra, looking for the lines of the high excitation in the ultraviolet (UV), which are not currently available. Future UV missions such as Spektr-UV \citep{2016ARep...60....1B} may help to solve the wind question. Diffuse X-ray emission, which also traces stellar wind in \hii{} regions \citep[e.~g.][]{2019Natur.565..618P, 2021SciA....7.9511L} has not been investigated in S255-257. Therefore, the question about the stellar wind remains for future studies.

The Photo-dissociation regions (PDRs) surrounding the S254-258 complex are irradiated by a moderate UV~field $G_0 < 100$~Habings, see \citet{Kirsanova_KGO}, comparable with the Horsehead and S235~PDRs \citep{Philipp_2006, Kirsanova_2021}. Using simulations of an expanding \hii{} region for a star with effective temperatures of $30\times10^3$~K \citep{Kirsanova_2020} we found values of $G_0$ and compared them with the observed ones in S255 and S257. A PDR with $G_0 > 20$~Habings spreads far over 2~pc from the ionizing stars for $n_{\rm init} = 10^2$~cm$^{-3}$. Therefore, this model does not work for S255 and S257. We found models with $n_{\rm init} = 10^3$~and~$10^4$~cm$^{-3}$ are more relevant for these regions as they produce PDRs with $5< G_0 < 300$~Habings.

These indications about the non-uniform gas distribution are also supported from simulations. \citet{Kirsanova_KGO} found that the widths of the PDRs in S255 and S257 range from 0.1 up to 0.5~pc for different values of the uniformly distributed gas density in the range $10^2 \leq n_{\rm init} \leq 10^4$~cm$^{-3}$, and these widths are too thin compared with the observed values of 0.3-0.4~pc. Simulations with $n_{\rm init} = 10^3-10^4$~cm$^{-3}$ also give \nelec{} values of 50-70~cm$^{-3}$, which agrees with our observations within a factor of 2. The regions with the high \nelec{} on the borders of the simulated \hii{} regions have widths $\leq 0.05$~pc, which is 3-5 times less than those observed. All these discrepancies can be overcome considering a non-uniform clumpy medium where expansion of the \hii{} regions takes place.

In spite of the importance of the clumpy and non-uniform medium for the study of the three-dimensional structure, another property of the absorbing material along the line of the sight, namely the enhancement of the total-to-selective extinction up to $R_V = 4$, can be explained by numerical models with uniformly distributed material. Destruction of small dust grains by UV~photons, or dust drift under radiation pressure in \hii{} regions, results in a decrease of the small dust grain abundance, as was found observationally by \citet{Marconi_1998, Witt_2006}, and subsequently shown by the simulations of \citet{2013ARep...57..573P, 2015MNRAS.449..440A}. 

\section{Conclusions}

We present observations of the star-forming complex S254-258 using the optical tunable-filter photometer MaNGaL on the Zeiss-1000 telescope at SAO RAS. Surface photometry in \Halpha, \Hbeta, \NII, \SII{} and \OIII{} lines is complemented by long-slit spectroscopic observations using SCORPIO-2 at the 6-m BTA telescope at SAO RAS. Our main conclusions are as follows.

\begin{itemize}
 
    \item The two extended \hii{} regions in the complex, S255 an S257, both have a more or less round shape in the plane of the sky. Both regions are attenuated by absorbing material with $2 \leq A_V \leq 5$~mag, with clear enhancements toward the molecular cloud located between them. The electron density in these regions rises from 100~cm$^{-3}$ near the ionizing stars, to 400~cm$^{-3}$ at the edge of this dense molecular cloud. Moreover, there is another enhancement of the electron density toward the outskirts of S255 and S257. These enhancements may be related to diffuse UV~photons which penetrate through the clumpy dense neutral material and ionize it. Another possibility is the evacuation of ionized gas from the vicinity of the massive stars by stellar winds. In order to test the latter suggestion, additional observations in X-rays or UV wavelengths are needed.
    
    \item The three-dimensional structure of S255 differs from that of S257. Specifically, S255 is surrounded by dense neutral gas from all sides, while S257 is situated on the border of a molecular cloud, and does not have dense front and rear walls. We propose that S257 represents a blister-type \hii{} region, as the density of both ionized and neutral gas components decreases away from the molecular cloud.
    
    \item There are two compact \hii{} regions, S256 and S258, which are deeply embedded into molecular clouds, with $A_V \approx 5$ and 10~mag, respectively. S256 probably represents a blister-type \hii{} region, but at an earlier stage in comparison to S257. The electron density in S256 decreases from 100~cm$^{-3}$ to 50~cm$^{-3}$, along with the amount of neutral material at both the front and rear walls of the \hii{} region. S258 has the highest electron density of all the regions ($600$~cm$^{-3}$), although we were not able to study the three-dimensional structure of this region due to the weakness of the optical lines.
    
    \item We suggest that the total-to-selective extinction $R_V$ towards these \hii{} regions is higher than in the diffuse ISM, due to destruction of small dust grains by UV photons or dust drift from radiation pressure. 

\end{itemize}

\section*{Acknowledgements}

We thank P.~M.~Zemlyanukha and I.~I.~Zinchenko for fruitful discussions of the star-forming complex. We are also thankful to the unknown referee for his/her very relevant comments.

This study was supported by the Russian Science Foundation, grant 21-12-00373. Observations with the SAO RAS telescopes are supported by the Ministry of Science and Higher Education of the Russian Federation. The renovation of telescope equipment is currently provided within the national project `Science and Universities'. 
\section*{Data Availability}

The data underlying this article will be shared on reasonable request to the corresponding author.



\bibliographystyle{mnras}
\bibliography{mnras_S255} 



\appendix

\section{Parameters of the star-forming S254-258 complex}\label{sec:appendixA}

\begin{figure*}
    \centering
    \includegraphics[width=2.0\columnwidth]{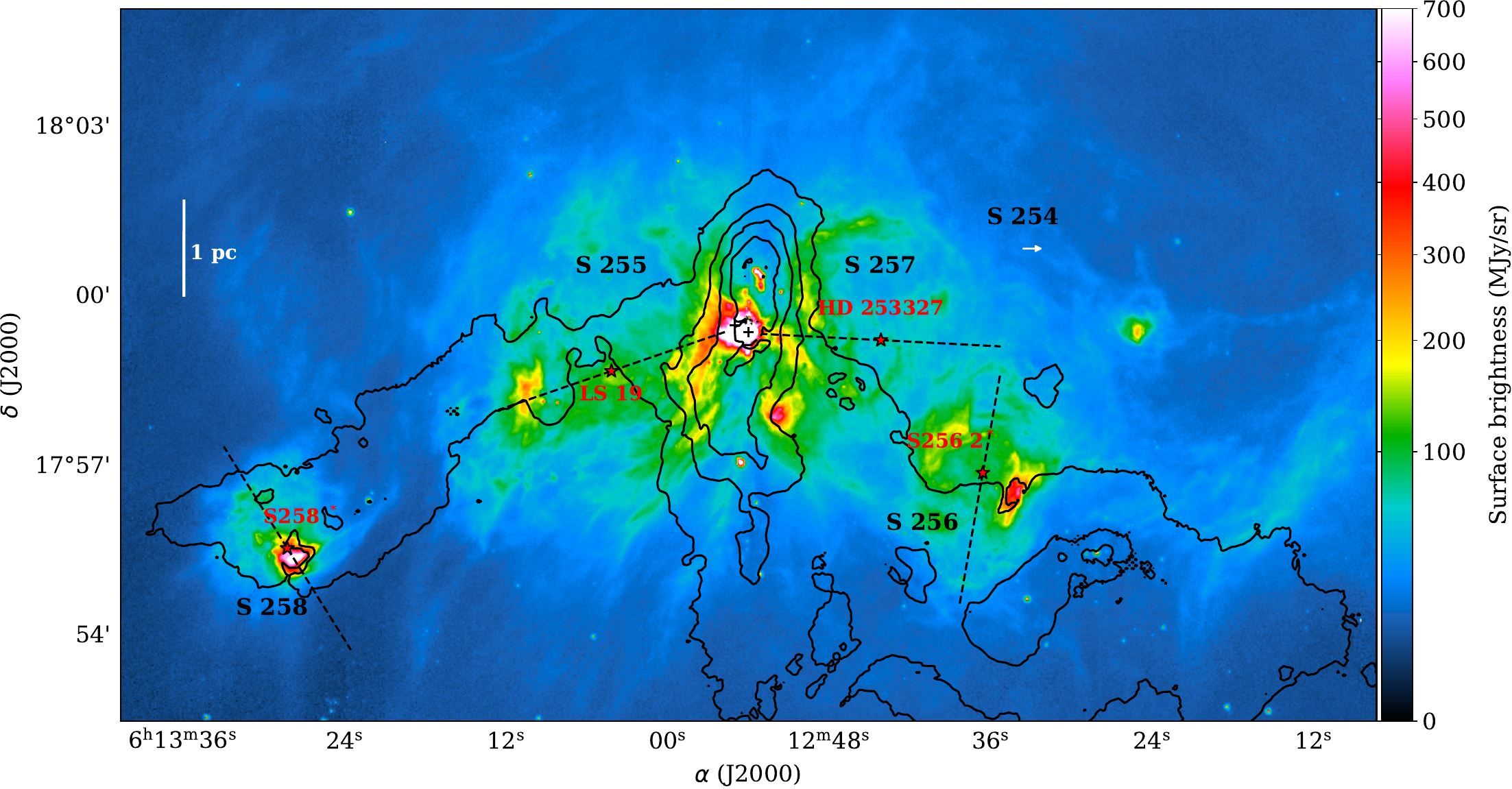}
    \caption{Mid-IR color image made by {\it Spitzer} telescope at 8\micron{} of the star-forming complex S254-258 with the Spitzer Super-Mosaic Pipeline (Digital Object Identifier (DOI): 10.26131/IRSA433). The black contours show hydrogen column density $N$(\ion{H}{I}+H$_2$) calculated with the {\it Herschel} data the 160-500\micron{} range by \citet{Ladeyschikov_2021}. The contours are given for 6, 12, 24 and $48\times 10^{20}$~cm$^{-2}$. The black dashed lines show the long slits for the optical spectroscopy in the present study. Red star with black border show the positions of the ionizing stars. A physical scale bar is shown by white line. The plus symbols show positions if infrared sources IRS~1 and IRS~3.}
    \label{fig:large_scale}
\end{figure*}

\begin{table}
    \centering
    \begin{tabular}{c|c|c}
    \hline
       \hii{} region  & Exciting star          & $T_e$\\
                      &                        &    (K)     \\
    \hline
        S254          &  O9.0V$^1$, O9.0V$^3$  &            \\
        S255          &  B0.0V$^1$, B0.0V$^3$  &  8200$^4$, 8833$\pm$170$^7$  \\
        S256          &  B2.5V$^2$, B0.9V$^3$  &               \\
        S257          &  B0.5V$^1$             &  6900-7900$^5$, 7970$\pm260^6$ \\
        S258          &  B3V$^2$, B1.5V$^3$    &                \\
       \hline
         & 
    \end{tabular}
    \caption{Parameters of the exciting stars and \hii{} regions from the literature used in the present study. References: $^1$\citet{Moffat_1979}, $^2$\citet{Russeil_2007}, $^3$\citet{Chavarria_2008}, $^4$\citet{Fern_Mart_2017}, $^5$\citet{Mendez_Delg_2022}, $^6$\citet{Esteban_2018}, $^7$G192.638-00.008 in \citet{Wenger_2019}.}
    \label{tab:HII_pars_lit}
\end{table}

\newpage
\section{Integrated spectra of the nebula}\label{sec:appendixB}

\begin{figure*}
    \centering
    \includegraphics[width=1.9\columnwidth]{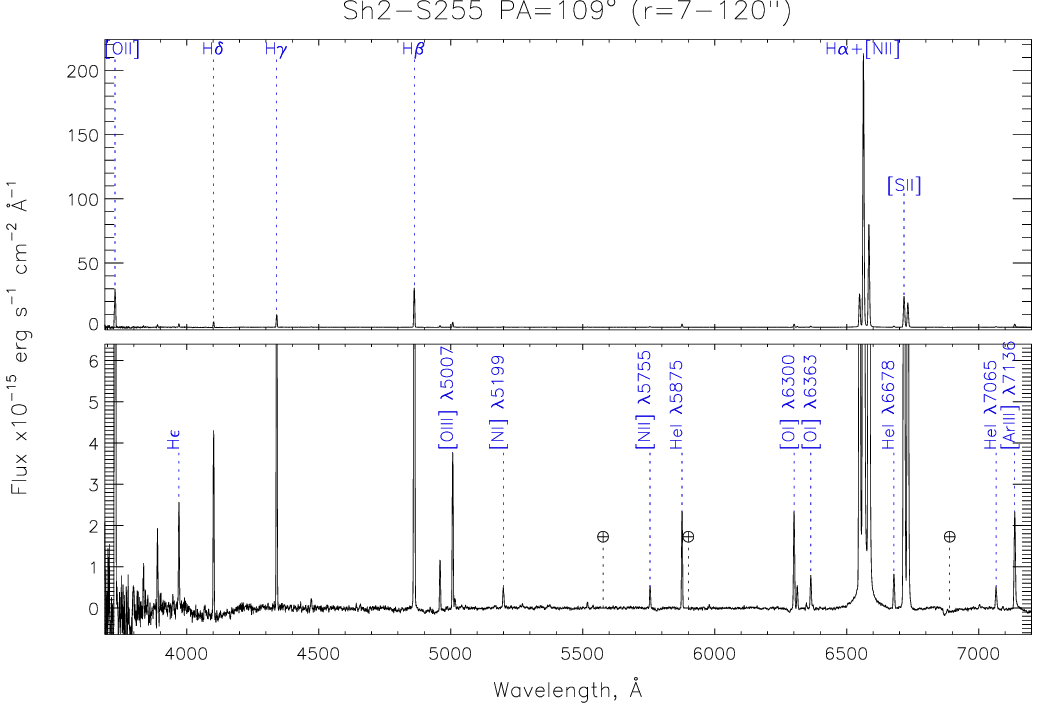}\\
        \includegraphics[width=1.9\columnwidth]{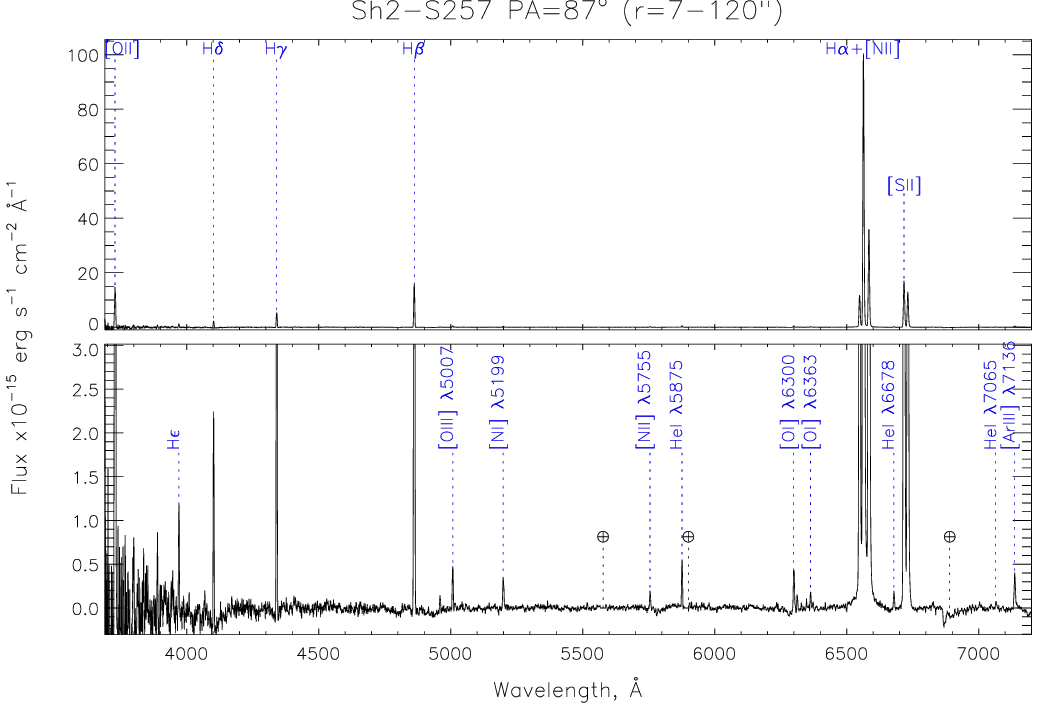}
 \caption{The integrated  continuum-subtracted optical spectra of S255 and S257. The integration range is shown at the top. The main object's  and telluric   emission lines are labelled in blue and black  respectively.  }       
    \label{fig:spec1}
\end{figure*}

\begin{figure*}
    \centering
    \includegraphics[width=1.9\columnwidth]{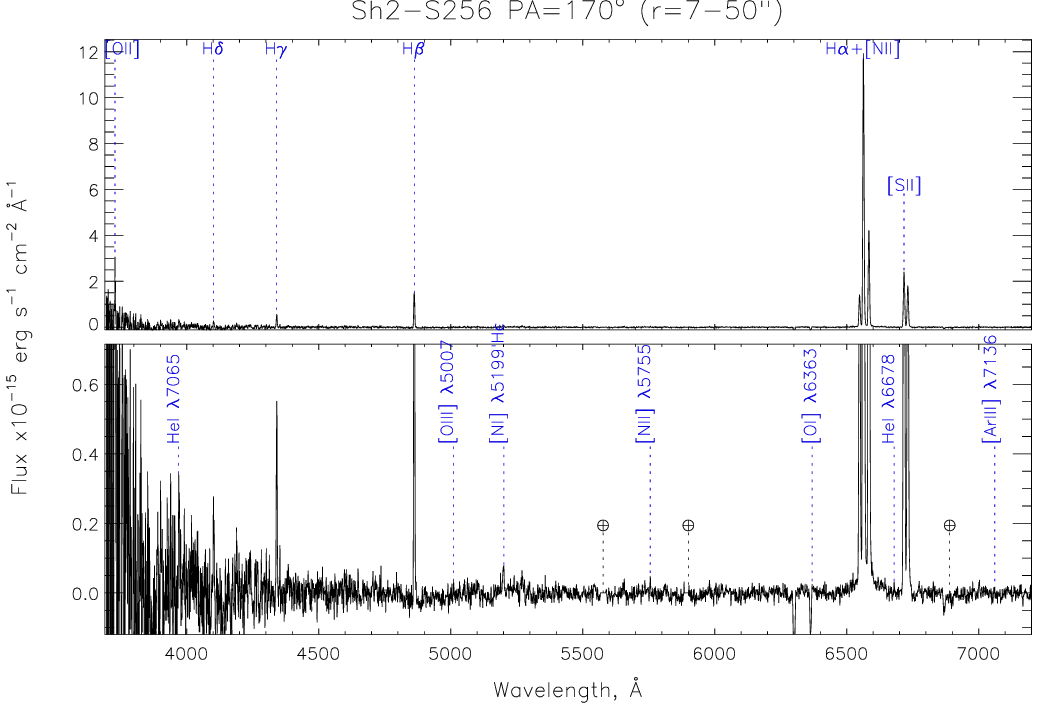}\\
        \includegraphics[width=1.9\columnwidth]{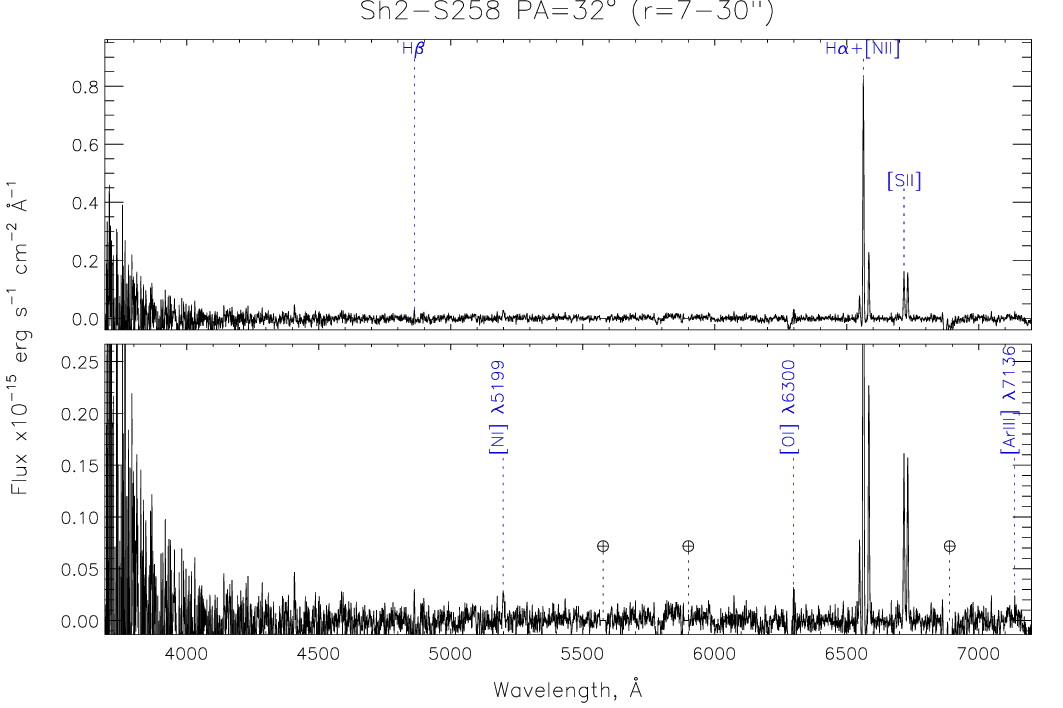}
    \caption{Same as Fig.~\ref{fig:spec1}, but for S256 and  S258.}
    \label{fig:spec2}
\end{figure*}

Fig.~\ref{fig:spec1} -- \ref{fig:spec2} present the integrated of all four regions observed with SCORPIO-2 in the long-slit mode after subtraction of continuum interpolated  by cubic spline. The central zone of the star formation regions ($r<7$ arcsec) was ignored during flux integration to avoid bright central star contamination in the total spectrum. The Poisson noise  grows significantly in the blue part of the spectra because  the total quantum efficiency of the used CCD and grism decreases at $\lambda<5000$\AA.  

The normalized intensities of the emission lines (Gaussian fitting results) are presented in Table~\ref{tab:linefluxes}. For two largest regions S255 and S257 we present results separately for the inner ($r<60$ arcsec) and outer ($r=60-120$ arcsec) radial range, because the relative intensity of high-excitation lines (\OIII, \HeI, \ArIII) is significantly lower in the outer regions. Table~\ref{tab:linefluxes_dered} presents the similar data after interstellar extinction corrected by using \Halpha/\Hbeta{} ratio in the integrated spectra. The corresponding \AV{} and $T_e$ values are also shown. All errors are $3\sigma$.

\begin{table*}
    \centering
    \begin{tabular}{r|r|r|r|r|r|r|r}
    \hline
        Line                    & \multicolumn{2}{c|}{Sh2-S255}    & Sh2-S256  &      \multicolumn{2}{c|}{Sh2-S257}   & Sh2-S258  \\
                           & ($r=7-60''$)    &  ($r=60-120''$) & ($r=7-50''$)  &   ($r=7-40''$)&  ($r=40-120''$)  &  ($r=7-30''$)  \\
         \OII~$\lambda$3727&  97.8$\pm$2.41  & 122.9 $\pm$3.20 & 133.3$\pm$14.8& 95.23$\pm$4.57&   107.7$\pm$4.66 &   --           \\
H$\varepsilon\,\lambda$3970&   7.01$\pm$0.32 &   8.54$\pm$0.39 &  20.0$\pm$3.51&  5.89$\pm$0.58&    6.03$\pm$0.55 &   --           \\
     H$\delta\,\lambda$4102&  12.14$\pm$0.27 &  13.11$\pm$0.33 &  15.4$\pm$2.44& 11.93$\pm$0.48&   10.90$\pm$0.50 &   --           \\ 
     H$\gamma\,\lambda$4340&  29.33$\pm$0.27 &  31.64$\pm$0.34 &  31.7$\pm$1.87& 31.64$\pm$0.49&   30.89$\pm$0.49 &   --           \\
      H$\beta\,\lambda$4861& 100.00$\pm$0.27 & 100.00$\pm$0.32 & 100.0$\pm$1.64&100.00$\pm$0.45&  100.00$\pm$0.44 & 100.00$\pm$22.0\\
        \OIII~$\lambda$5007&  19.41$\pm$0.15 &   2.65$\pm$0.12 &  --           &  5.73$\pm$0.22&    0.40$\pm$0.18 &   --           \\
          \NI~$\lambda$5199&   1.20$\pm$0.09 &   2.77$\pm$0.13 &   7.4$\pm$1.52&  2.08$\pm$0.22&    2.99$\pm$0.26 & 130.92$\pm$40.5\\ 
         \NII~$\lambda$5755&   1.54$\pm$0.08 &   1.89$\pm$0.09 &   1.8$\pm$0.50&  1.15$\pm$0.15&    1.12$\pm$0.15 &   --           \\
         \HeI~$\lambda$5876&  12.62$\pm$0.11 &   1.44$\pm$0.09 &  --           &  6.24$\pm$0.17&    0.79$\pm$0.16 &   --           \\
          \OI~$\lambda$6300&   6.21$\pm$0.14 &  11.79$\pm$0.21 &  --           &  2.28$\pm$0.27&    3.99$\pm$0.34 &   --           \\
        \SIII~$\lambda$6312&   2.07$\pm$0.09 &   1.65$\pm$0.13 &  --           &  0.88$\pm$0.17&    0.78$\pm$0.18 &   --           \\
          NII~$\lambda$6548& 101.18$\pm$0.36 & 111.28$\pm$0.46 & 119.9$\pm$1.46& 85.72$\pm$0.51&   94.27$\pm$0.55 & 200.60$\pm$20.8\\
     H$\alpha\,\lambda$6563& 893.23$\pm$2.53 & 778.28$\pm$2.60 & 987.6$\pm$3.16&755.87$\pm$3.49&  753.80$\pm$3.46 &2705.08$\pm$39.2\\
         \NII~$\lambda$6583& 309.35$\pm$0.93 & 341.51$\pm$1.20 & 356.0$\pm$2.05&259.84$\pm$1.28&  286.26$\pm$1.39 & 738.48$\pm$28.0\\
         \HeI~$\lambda$6678&   5.43$\pm$0.10 &   0.37$\pm$0.09 &   1.0$\pm$0.36&  2.26$\pm$0.15&    0.20$\pm$0.10 &   --           \\
         \SII~$\lambda$6716&  80.69$\pm$0.30 & 127.57$\pm$0.51 & 202.4$\pm$1.69&106.97$\pm$0.61&  156.50$\pm$0.82 & 527.75$\pm$26.2\\
         \SII~$\lambda$6731&  64.19$\pm$0.26 & 100.70$\pm$0.43 & 155.9$\pm$1.58& 82.49$\pm$0.50&  119.19$\pm$0.66 & 526.03$\pm$26.3\\
         \HeI~$\lambda$7065&   3.39$\pm$0.10 &   0.16$\pm$0.05 &  --           &  1.16$\pm$0.18&    --            &   --           \\
       \ArIII~$\lambda$7136&  15.50$\pm$0.13 &   1.77$\pm$0.11 &  --           &  5.58$\pm$0.19&    --            &  30.92$\pm$9.27\\

       \hline
    \end{tabular}
    \caption{Relative intensities of emission lines in the integrated spectra of S255-S258 regions (I(\Hbeta) = 100) before reddening corrections.  Flux errors are 3$\sigma$.}
    \label{tab:linefluxes}
\end{table*}

\begin{table*}
    \centering
    \begin{tabular}{r|r|r|r|r|r|r|r}
    \hline
        Line                    & \multicolumn{2}{c|}{Sh2-S255}    & Sh2-S256  &      \multicolumn{2}{c|}{Sh2-S257}   & Sh2-S258  \\
                           &($r=7-60''$)    &  ($r=60-120''$)& ($r=7-50''$)   &   ($r=7-40''$) &  ($r=40-120''$)&  ($r=7-30''$)  \\
         \OII~$\lambda$3727&246.28$\pm$ 6.07&275.85$\pm$ 7.18&364.78$\pm$40.47&208.46$\pm$10.00&235.34$\pm$10.18&    --          &\\
H$\varepsilon\,\lambda$3970& 14.47$\pm$ 0.66& 16.12$\pm$ 0.74& 44.28$\pm$ 7.76& 10.92$\pm$ 1.08& 11.13$\pm$ 1.02&    --          &\\
     H$\delta\,\lambda$4102& 22.61$\pm$ 0.50& 22.58$\pm$ 0.57& 30.43$\pm$ 4.79& 20.21$\pm$ 0.81& 18.44$\pm$ 0.85&    --          &\\
     H$\gamma\,\lambda$4340& 45.43$\pm$ 0.42& 46.40$\pm$ 0.50& 51.16$\pm$ 3.01& 45.88$\pm$ 0.71& 44.74$\pm$ 0.71&    --          &\\
      H$\beta\,\lambda$4861&100.00$\pm$ 0.27&100.00$\pm$ 0.32&100.00$\pm$ 1.64&100.00$\pm$ 0.45&100.00$\pm$ 0.44&100.00$\pm$22.04&\\
        \OIII~$\lambda$5007& 17.07$\pm$ 0.13&  2.37$\pm$ 0.11&    --          &  5.14$\pm$ 0.20&  0.35$\pm$ 0.16&    --          &\\
          \NI~$\lambda$5199&  0.90$\pm$ 0.07&  2.15$\pm$ 0.10&  5.43$\pm$ 1.11&  1.64$\pm$ 0.17&  2.35$\pm$ 0.20& 74.80$\pm$23.17&\\
         \NII~$\lambda$5755&  0.79$\pm$ 0.04&  1.05$\pm$ 0.05&  0.87$\pm$ 0.23&  0.65$\pm$ 0.08&  0.64$\pm$ 0.09&    --          &\\
         \HeI~$\lambda$5876&  5.99$\pm$ 0.05&  0.75$\pm$ 0.05&    --          &  3.32$\pm$ 0.09&  0.42$\pm$ 0.09&    --          &\\
          \OI~$\lambda$6300&  2.34$\pm$ 0.05&  5.02$\pm$ 0.09&    --          &  1.00$\pm$ 0.12&  1.74$\pm$ 0.15&    --          &\\
        \SIII~$\lambda$6312&  0.78$\pm$ 0.03&  0.69$\pm$ 0.05&    --          &  0.39$\pm$ 0.07&  0.34$\pm$ 0.08&    --          &\\
          NII~$\lambda$6548& 33.71$\pm$ 0.12& 42.51$\pm$ 0.18& 36.17$\pm$ 0.44& 33.72$\pm$ 0.20& 37.18$\pm$ 0.22& 22.89$\pm$ 2.38&\\
     H$\alpha\,\lambda$6563&295.49$\pm$ 0.84&295.48$\pm$ 0.99&295.45$\pm$ 0.95&295.47$\pm$ 1.36&295.49$\pm$ 1.36&302.83$\pm$ 4.37&\\
         \NII~$\lambda$6583&101.36$\pm$ 0.30&128.58$\pm$ 0.45&105.38$\pm$ 0.61&100.75$\pm$ 0.50&111.31$\pm$ 0.54& 80.70$\pm$ 3.06&\\
         \HeI~$\lambda$6678&  1.70$\pm$ 0.03&  0.13$\pm$ 0.03&  0.28$\pm$ 0.10&  0.85$\pm$ 0.06&  0.08$\pm$ 0.04&    --          &\\
         \SII~$\lambda$6716& 24.84$\pm$ 0.09& 45.49$\pm$ 0.18& 55.99$\pm$ 0.47& 39.35$\pm$ 0.22& 57.74$\pm$ 0.30& 51.07$\pm$ 2.54&\\
         \SII~$\lambda$6731& 19.63$\pm$ 0.08& 35.69$\pm$ 0.15& 42.81$\pm$ 0.43& 30.17$\pm$ 0.18& 43.71$\pm$ 0.24& 50.38$\pm$ 2.52&\\
         \HeI~$\lambda$7065&  0.90$\pm$ 0.03&  0.05$\pm$ 0.02&    --          &  0.37$\pm$ 0.06&    --          &    --          &\\
       \ArIII~$\lambda$7136&  3.98$\pm$ 0.03&  0.54$\pm$ 0.03&    --          &  1.76$\pm$ 0.06&    --          &  2.19$\pm$ 0.66&\\
    \hline
A$_V$           &   2.96 $\pm$0.01&  2.81 $\pm$0.01 & 3.23 $\pm$0.01& 2.51 $\pm$0.01 & 2.51 $\pm$0.01 & -- \\
$T_e$(\NII), K  &   8089 $\pm$133 &  8218 $\pm$130 & 8222 $\pm$ 762 & 7615 $\pm$290 & 7357 $\pm$311 & -- \\
    
       \hline
    \end{tabular}
    \caption{Reddening-corrected relative intensities of emission lines in the integrated spectra of S255-S258 regions (I(\Hbeta) = 100).  Flux errors are 3$\sigma$.}
    \label{tab:linefluxes_dered}
\end{table*}

\newpage
\section{Spectra and SEDs of the ionizing stars}\label{sec:appendixC}

\begin{figure*}
    \centering
    \includegraphics[width=1.9\columnwidth]{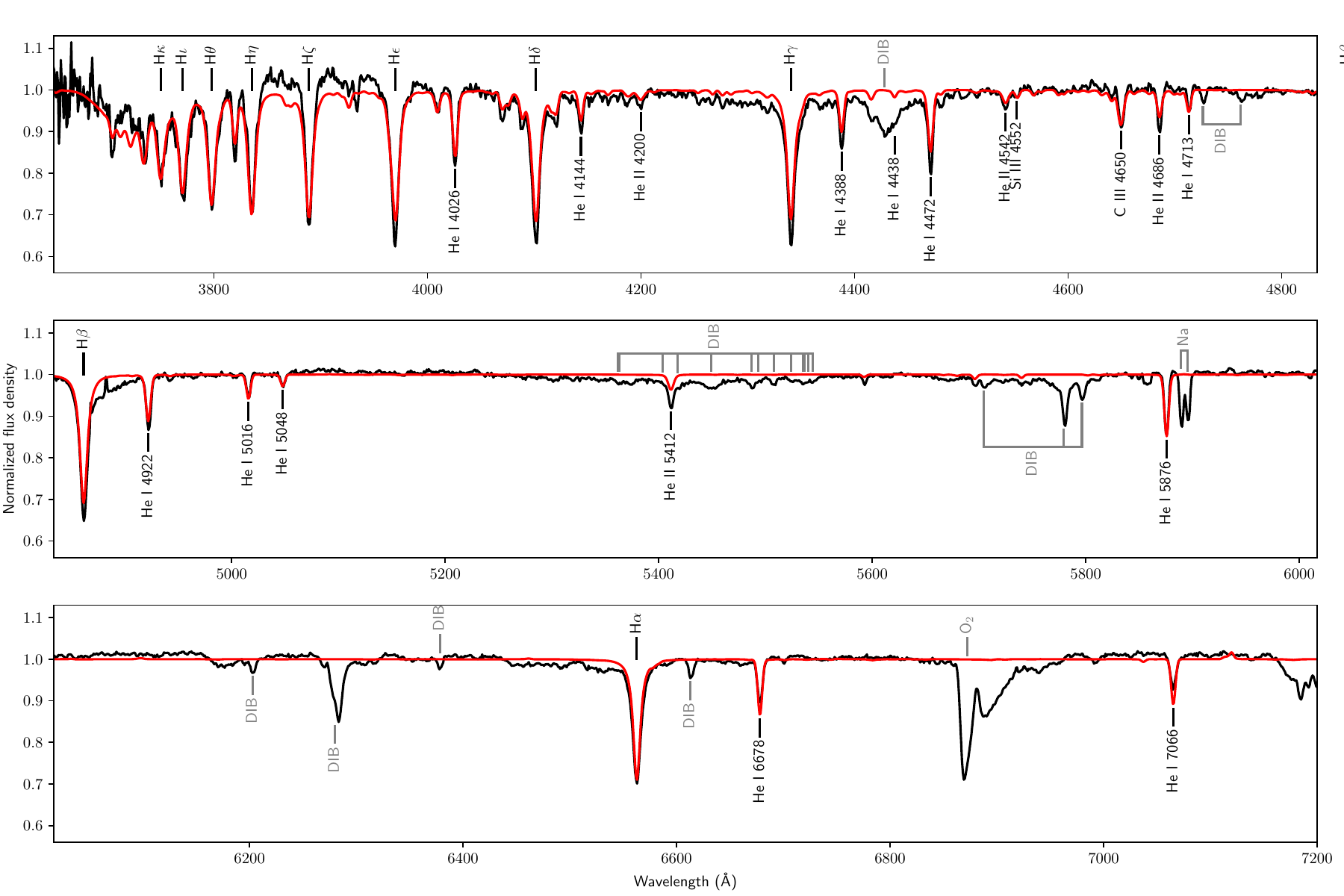}
    \caption{Normalized spectrum of the ionizing star of S255 (black) together with the best-fit stellar atmosphere (red). Photospheric lines are labeled in black, while interstellar and telluric lines are labeled in gray.}
    \label{fig:S255_stellar_spect}
\end{figure*}

\begin{figure*}
    \centering
    \includegraphics[width=1.9\columnwidth]{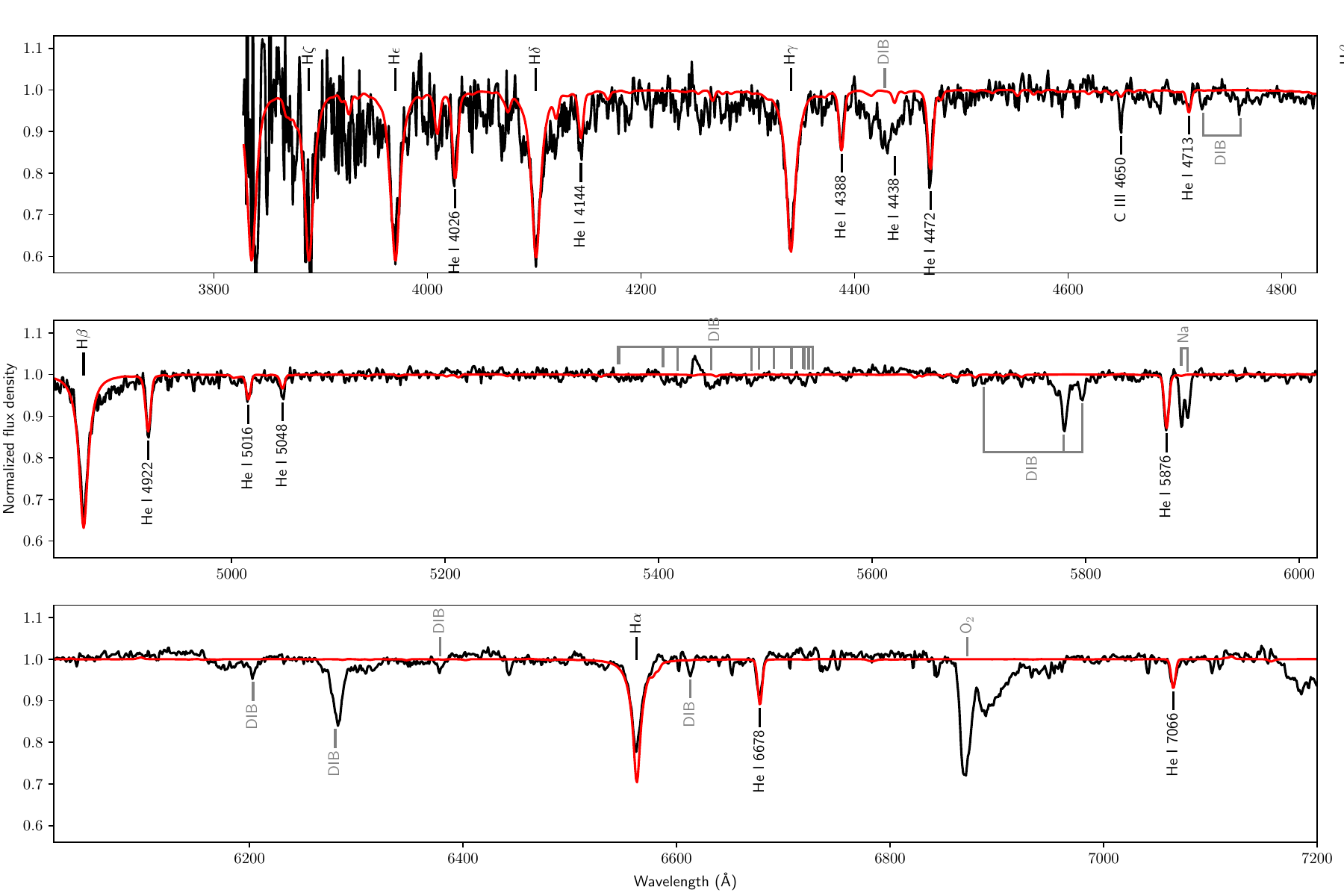}
    \caption{Same as Fig.~\ref{fig:S255_stellar_spect}, but for S256.}
    \label{fig:S256_stellar_spect}
\end{figure*}

\begin{figure*}
    \centering
    \includegraphics[width=1.9\columnwidth]{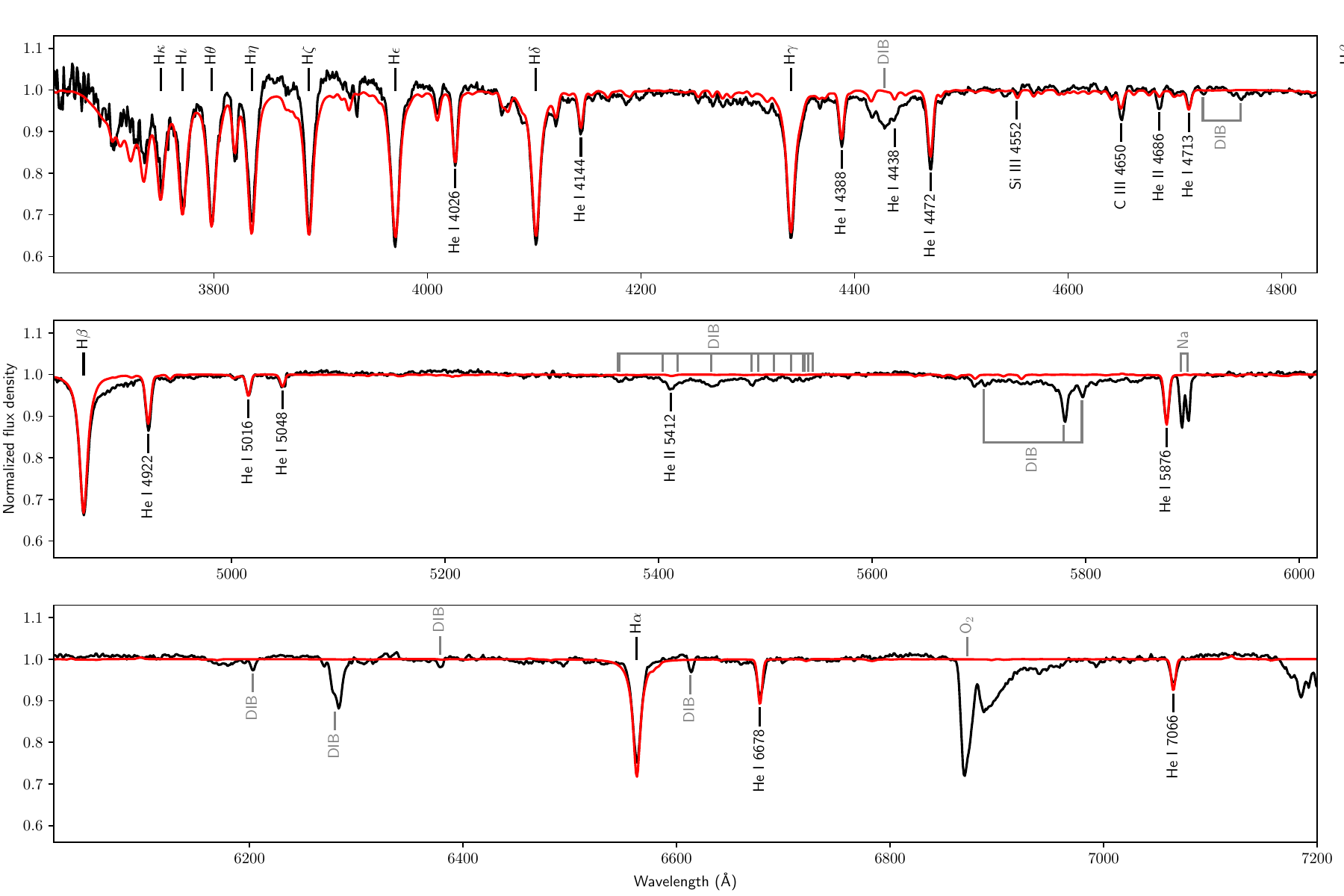}
    \caption{Same as Fig.~\ref{fig:S255_stellar_spect}, but for S257.}
    \label{fig:S257_stellar_spect}
\end{figure*}

\begin{figure}
    \centering
    \includegraphics[width=\columnwidth]{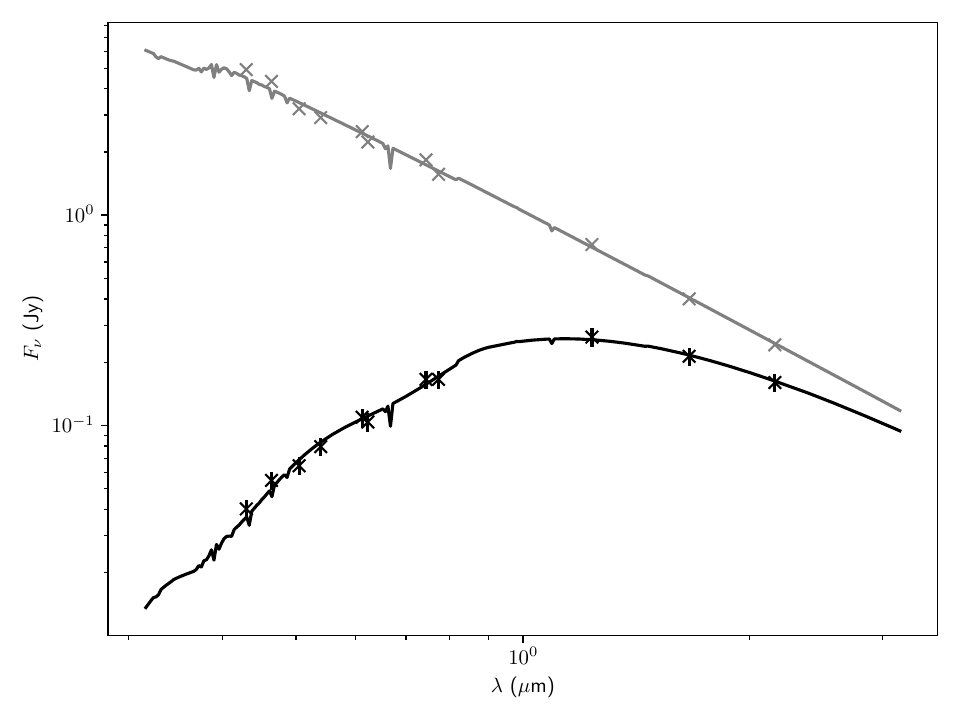}
    \caption{Fit of the observed (reddened) SED of the ionizing star of S255 (black), together with the dereddened spectrum (gray). Photometric measurements are shown crosses, with the measurement error indicated for the observed values.}
    \label{fig:S255_SED}
\end{figure}

\begin{figure}
    \centering
    \includegraphics[width=\columnwidth]{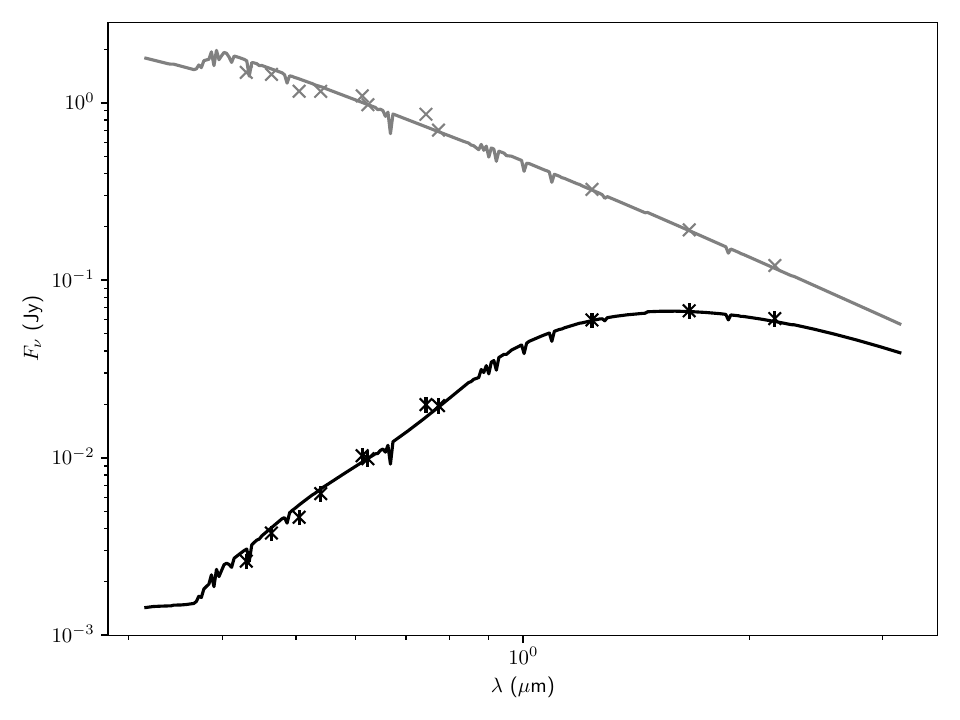}
    \caption{Same as Fig.~\ref{fig:S256_SED}, but for S256.}
    \label{fig:S256_SED}
\end{figure}

\begin{figure}
    \centering
    \includegraphics[width=\columnwidth]{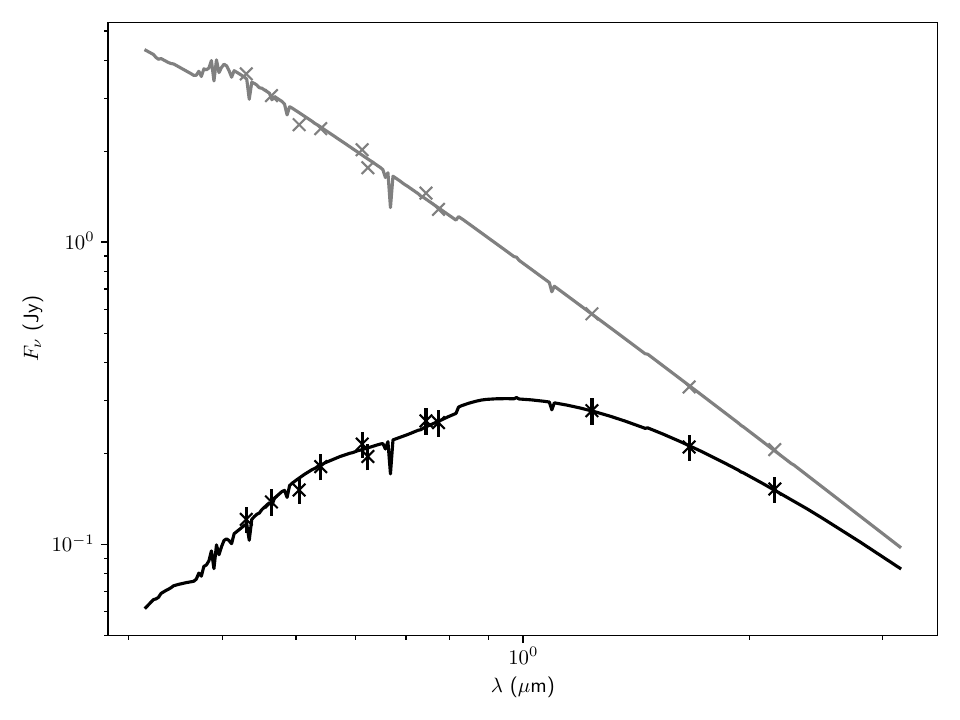}
    \caption{Same as Fig.~\ref{fig:S257_SED}, but for S257.}
    \label{fig:S257_SED}
\end{figure}


\bsp	
\label{lastpage}
\end{document}